# Improving Musical Instrument Classification with Advanced Machine Learning Techniques
## A Comparative Study

**Joanikij Chulev**
joanikijchulev@proton.me

Utrecht University

May 2024

## Abstract

*Musical instrument classification, a key area in Music Information Retrieval, has gained considerable interest due to its applications in education, digital music production, and consumer media. Recent advances in machine learning, specifically deep learning, have enhanced the capability to identify and classify musical instruments from audio signals. This study applies various machine learning methods, including Naive Bayes, Support Vector Machines, Random Forests, Boosting techniques like AdaBoost and XGBoost, as well as deep learning models such as Convolutional Neural Networks and Artificial Neural Networks. The effectiveness of these methods is evaluated on the NSynth dataset, a large repository of annotated musical sounds. By comparing these approaches, the analysis aims to showcase the advantages and limitations of each method, providing guidance for developing more accurate and efficient classification systems. Additionally, hybrid model testing and discussion are included. This research aims to support further studies in instrument classification by proposing new approaches and future research directions.*

# 1    Introduction

Research in image data has been quite extensive, particularly fueled by the advent of large-scale datasets like ImageNet, which has driven advancements in computer vision. The popularity of social media platforms has also significantly boosted interest in image data, as images become central in consumer content, communication, marketing, and media. Image content analysis, including object recognition and image retrieval, has seen rapid growth due to its direct applications in digital libraries, autonomous driving, and surveillance systems [1]. Similarly, text data has been at the forefront of many AI breakthroughs, particularly with natural language processing (NLP) technologies. The ubiquity of text in web content, digital communications, and business documents makes it a primary focus for developments in algorithms that can parse, understand, and generate text. Meanwhile, machine translation has revolutionized communication by breaking language barriers, making information more accessible globally. Automated summarization tools also vastly help in digesting large volumes of information quickly, essential in today's world filled with abundance of information [2].

In contrast, audio data has not seen the same level of extensive research or application, although interest is growing. Historically, the complexity of sound and its temporal nature make it a more challenging domain. Unlike images, which are static and have well-defined spatial relationships, or text, which has a clear structure, audio requires dealing with time-dependent signals that often contain a mixture of sounds that must be separated and identified. However, projects like Audio Set aim to bridge this gap by providing a large-scale dataset annotated with a wide range of sound events, which can help stimulate the development of more sophisticated audio processing technologies [3]. Thus, further research work in audio analysis needs to be implemented.

Audio analysis plays a critical role in improving communication and music media technologies, such as speech recognition systems, which are essential for creating more effective human-computer interactions [4]. Audio analysis can enhance security measures by detecting anomalous sounds, like glass breaking or unexpected entry or gun shots, which are crucial for surveillance and security systems [5]. All these applications have become more refined with the introduction of Artificial Intelligence (AI). It significantly enhances the capability of audio analysis through sophisticated algorithms that can learn from and adapt to various audio environment uses. AI techniques, such as Machine Learning (ML) and Deep Learning (DL), are used to automatically and accurately classify, segment, and even generate audio signals based on learned patterns. AI algorithms can identify specific sounds from a mix, making it easier to focus on relevant audio cues in complex environments [4]. AI also facilitates interactive systems that respond to voice commands or adjust audio outputs in real-time according to the user preferences or environmental conditions [6].

Our main focus will be on musical applications and classification. In the media and entertainment industries, audio analysis plays a crucial role in enhancing user experiences, sound classification, and sound quality refinement. It enables more accurate metadata tagging, improved content recommendation systems, and advanced audio editing techniques, leading to superior auditory experiences and musical identification [7]. The classification of musical instruments using ML has emerged as a vibrant area of research within the fields of Music Information Retrieval (MIR) and AI. As the diversity and complexity of music grows, so does the challenge of accurately identifying and classifying musical instruments from audio signals. This task not only enhances our understanding of music but also serves various practical applications, such as music education, digital music production, and automated music transcription. The classification of musical instruments through ML involves extracting and learning from features such as timbre, pitch, and duration of notes to distinguish between different types of instruments [8].

Recent advancements in ML offer new potential for improving the accuracy of musical instrument classification. Techniques such as DL have shown remarkable success due to their ability to capture complex patterns in data without requiring manual feature engineering. Convolutional Neural Networks (CNNs), for instance, have been effectively applied to classify musical instruments by learning directly from the raw audio waveform or spectrogram representations [9]. This gives further insight into our comparison approach for model selection.

Furthermore, comparative studies provide valuable insights into the effectiveness of various ML models in musical instrument classification. By examining different approaches, researchers can identify the strengths and weaknesses of each technique, guiding the development of more robust and accurate classification systems. Such studies often explore a range of classifiers including Support Vector Machines (SVM), k-Nearest Neighbors (kNN), and more sophisticated neural network models, assessing their performance across different datasets and feature sets [10]. The continuous evolution of ML techniques offers significant potential for improving the accuracy and efficiency of musical instrument classification.

As a critical component of MIR, the precise identification of musical instruments in audio files is essential. This comparative study delves into these advancements by evaluating state-of-the-art algorithms, aiming to highlight the methods in this specific but rapidly growing field. In this study, multiple approaches to musical instrument classification are extensively tested. These approaches vary widely in their methodologies, computational requirements, and the underlying principles by which they classify audio data. Each method has been subjected to rigorous testing across various conditions. Based on these insights, further on we can propose several suggestions for future work.

## 2    Audio Data

Audio data refers to the representation and manipulation of the encoded sound information that computers and digital devices store and process [4]. Instrument sounds can be saved via this format. Various audio file types are used in ML to handle audio tasks such as classification, recognition, and analysis. Given the importance of data quality and structure in ML, assembling an appropriate dataset is vital for achieving high accuracy in tasks such as musical instrument classification.

For ML, the choice of audio file format can significantly impact the performance of audio processing models. High-quality, lossless formats like WAV or FLAC are generally best when the fidelity of the audio data is crucial for the task, such as in sound classification and music analysis tasks [11]. Furthermore, for ML, high-quality data can lead to more accurate models, especially in tasks requiring detailed audio analysis [12]. Thus, we would require a dataset that meets our criteria of multiple instrument classes, multiple sample sources and a high number of samples in high quality audio format.

The choice for the dataset and data manipulation of the dataset was also partially influenced by a thesis, acknowledging the results presented [13].

The NSynth [14] dataset was selected due to its comprehensive and high-quality collection of musical sounds, generated from a diverse array of instruments.

## 2.1    NSynth Dataset

The NSynth dataset is a collection of musical notes, specifically designed for the training and evaluation of ML models that deal with audio processing. Created by researchers at Google, NSynth stands for "Neural Synthesizer." The NSynth dataset comprises 305,979 musical notes, each distinguished by its individual pitch, timbre, and envelope characteristics. These notes are derived from 1,006 instruments sourced from commercial sample libraries. To create the dataset, four-second, monophonic audio snippets, termed "notes," were generated at a sampling rate of 16 kHz. The notes span the entire range of a standard MIDI piano (from 21 to 108) across five different velocities (25, 50, 75, 100, 127). During synthesis, the note sustains for the initial three seconds before undergoing a decay phase in the final second [14].

Due to instrument limitations, some instruments cannot produce all 88 pitches within the specified range, resulting in an average of 65.4 pitches per instrument [14]. This is due to the nature of the instruments, consequently it will pose no issue for the task of instrument classification. One of the key features of NSynth is that it includes both traditional musical instruments (acoustic) and electronic sounds (synthetic and electronic), providing a rich variety of timbres and sonic characteristics. The sources for the instruments are given in Table 1 below.

*Table 1: Instrument Sources.*

| Index | ID |
| --- | --- |
| **0** | acoustic |
| **1** | electronic |
| **2** | synthetic |

Instruments are also categorized based on high-level instrument families, of which there are 11. Each instrument (and all its notes) is labeled with exactly one. These families are given in Table 2 below.

*Table 2: Instrument Families.*

| Index | ID |
| --- | --- |
| **0** | bass |
| **1** | brass |
| **2** | flute |
| **3** | guitar |
| **4** | keyboard |
| **5** | mallet |
| **6** | organ |
| **7** | reed |
| **8** | string |
| **9** | synth_lead |
| **10** | vocal |

To facilitate ML model development, including training, validation, and testing phases, the NSynth dataset is divided into three distinct sets: Train Set - 289,205 samples, Valid Set - 12,678 samples, Test Set - 4,096 samples.

It is important to note that the instruments in the train set do not overlap with those in the validation or test sets, ensuring that the model's performance is evaluated on completely unseen data. The separation of instruments across the sets ensures that the evaluation metrics reflect the model's ability to generalize across different musical characteristics.

The NSynth dataset [14] is available for public download at: https://magenta.tensorflow.org/datasets/nsynth.

## 2.2 Wave File Features

To formulate a data frame that is understandable for a model we need to examine and extract features (characteristics) of the audio WAV files. Feature extraction from WAV files is a critical step in audio classification, as it involves translating raw audio data into a format that ML algorithms can process and analyze. WAV files store audio signals in a digital format, where sound is represented by a series of audio samples. To extract meaningful features from these files, various signal processing techniques are employed. Commonly extracted features include Mel-Frequency Cepstral Coefficients (MFCCs), which capture the timbre of the sound; spectral contrast, which details the difference in amplitude between peaks and valleys in the spectrum; and chroma features, which relate to the twelve different pitch classes. Additionally, temporal features like zero-crossing rate, which measures the rate at which the signal changes from positive to negative and vice versa, and energy, which represents the loudness, are also commonly extracted.

For our specific numerical feature dataset, we extracted the *Harmonic Percussive Index (HPI), Chroma Energy, Mel Spectrogram, MFCCs and Spectral Contrast* as features from the WAV files.

### 2.2.1 Harmonic Percussive Index

Percussive instruments are musical instruments that produce sound primarily by being struck, shaken, or scraped by a beater; hit against another similar instrument; or played by hand [15]. Examples include drums, cymbals, and xylophones. The sound is generated by the vibration of the instrument itself, without the need for strings or air columns. Percussive instruments are characterized by their ability to provide rhythm and are often referred to as the backbone of a musical ensemble's rhythm section.

Harmonic instruments, on the other hand, produce sound through the vibration of strings, air columns, or their electronic equivalents, and can produce harmonic tones. These instruments typically play a more melodic role in music. Examples include guitars, violins, flutes, and pianos. The sound produced by harmonic instruments is richer in overtones and can sustain notes, allowing for the creation of melody and harmony in music.

Harmonic instruments are typically smooth in time because their sound waves are continuous and change slowly. In contrast, percussive instruments are considered smooth in frequency since they produce brief, transient sounds that do not vary melodically but have a clear onset and decay. However, certain percussive instruments or samples, can produce longer sustained notes that have distinct pitches and harmonic-melodic qualities [16]. Conversely, harmonic instruments can sometimes produce percussive effects. For example, techniques such as pizzicato on string instruments, where the strings are plucked rather than bowed, generate a more percussive sound, emphasizing the attack much like a drum hit. Similarly, staccato playing on a violin or a trumpet can mimic the sharp, abrupt qualities typically associated with percussive sounds [16]. Based on these insights, we need to take the HPIs into account as it provides important information on the instrument type. Consult Figures 1, 2, 3 to see the harmonic and percussive elements visually displayed.

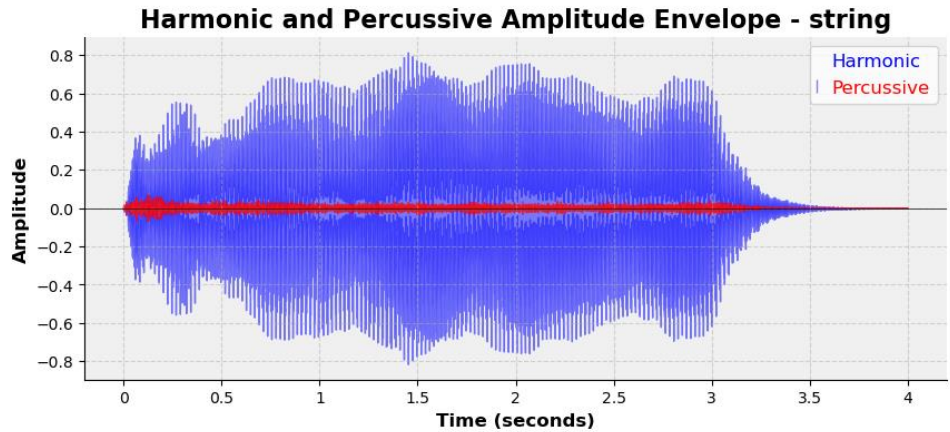

*Figure 1: Example of harmonic and percussive elements of an audio file, with superior harmonic elements.*

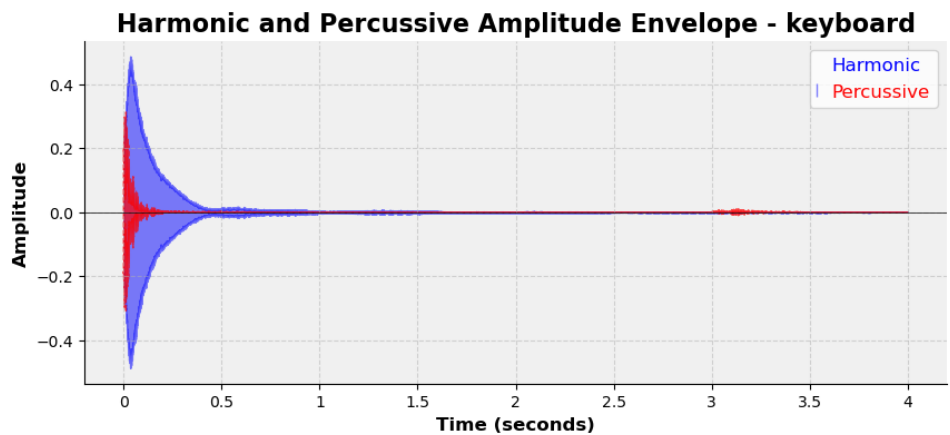

*Figure 2: Example of harmonic and percussive elements of an audio file, with similar harmonic and percussive elements.*

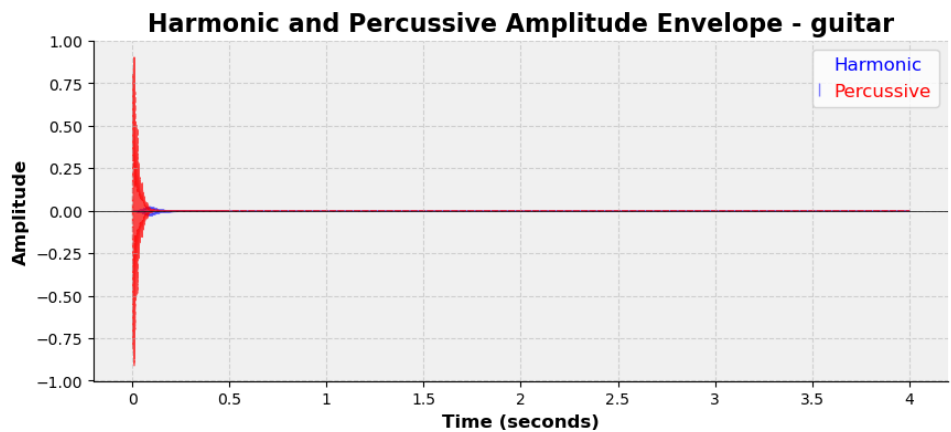

*Figure 3: Example of harmonic and percussive elements of an audio file, with superior percussive elements.*

### 2.2.2 Chroma Energy

Chroma features capture the essence of the twelve different pitch classes in Western music. This capability makes them invaluable for tasks that require the identification of harmonic and melodic elements in audio signals [17]. For instance, a study by Ghosal, Dutta, and Banerjee in 2018 utilized chroma-based features to effectively differentiate string instruments by analyzing the strength of notes in the Western 12-note scale extracted from audio signals. This method showcased how different string instruments leave distinct chroma signatures when producing sounds, which could be recognized and classified using ML classifiers like Neural Networks [18]. In musical instrument classification, chroma features help in distinguishing instruments based on their harmonic contributions, even when they produce similar pitches. For example, even if an instrument like a marimba and a string instrument like a cello or violin play the exact same note, their chroma features would differ due to the unique harmonic overtones each instrument produces, which are captured in the chroma energy representation [17]. Consult Figures 4, 5 below to see the visual representations of chroma energy and how they differ.

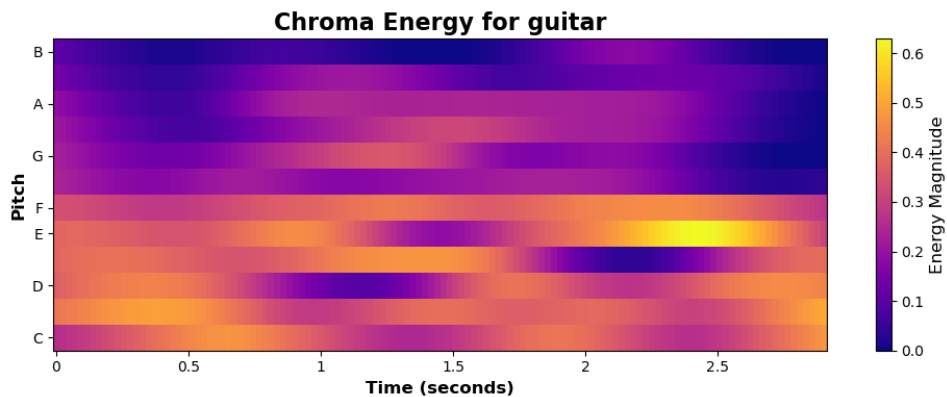

*Figure 4: Chroma energy representation of an audio sample.*

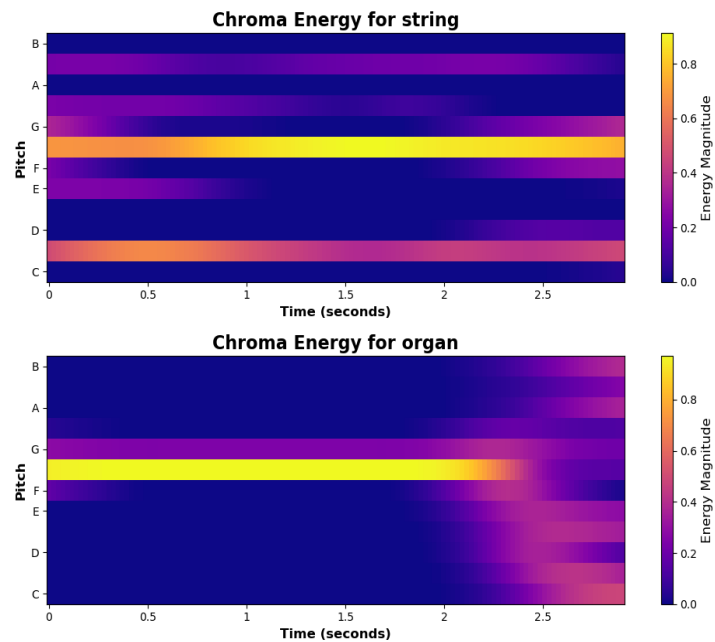

*Figure 5: Chroma energy representation of two audio samples played in the same pitch key – F Sharp.*

### 2.2.3    Mel Spectrogram

Mel spectrograms are a fundamental component in audio analysis, offering a quality method for capturing the textural nuances of audio signals. A Mel spectrogram is a visual representation of the spectrum of frequencies in a sound or music signal as they vary with time but adjusted to match the human ear's perception of sound scales. This adjustment involves using the Mel scale to emphasize perceptually important frequency bands and is particularly effective for identifying musical characteristics. For instance, a study by Agera et al. in 2015 explored the use of Mel spectrograms to extract textural features like Local Binary Patterns (LBP) for music genre classification, which can be similarly applied to instrument classification due to the textural similarities in both tasks. This approach allows classifiers, such as SVMs, to achieve high recognition rates by analyzing the unique textural patterns present in the Mel spectrogram of each instrument or music genre [19]. Comparative studies, such as the one conducted by Rodin et al. in 2023, have shown that Mel spectrograms, when used alongside DL architectures like VGG-16 and ResNet-34, provide a substantial improvement in the accuracy of musical instrument classification. These models leverage the detailed information encoded in the Mel spectrograms, proving more effective than basic audio features [20]. Although promising results were reached, instruments were classified into only three families. Nevertheless, this shows the importance of this feature in almost every application of audio analysis. Consult Figure 6 below to see the visual representations of Mel Spectrograms.

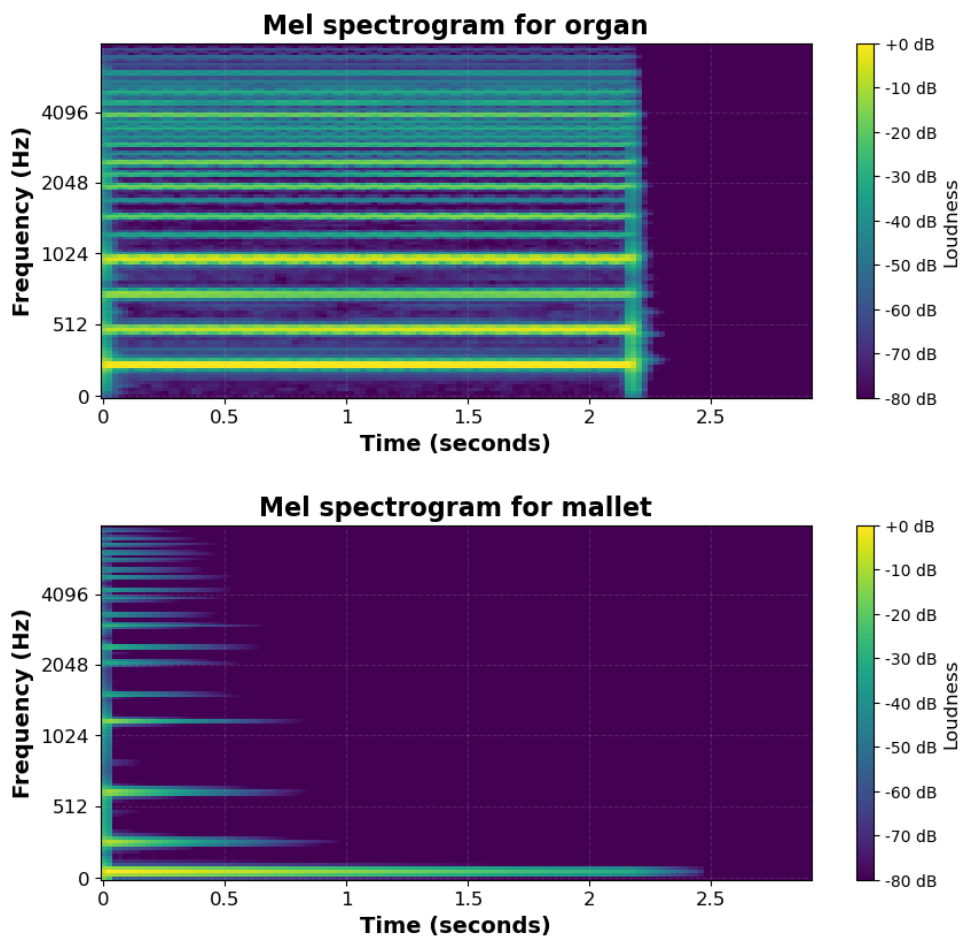

*Figure 6: Mel Spectrogram representations of two samples of instruments.*

### 2.2.4    MFCCs

MFCCs are a widely used feature in the classification of musical instruments due to their effectiveness. MFCCs are derived from the logarithm of the spectrum obtained through a Mel-scale filter bank, which closely mimics the human auditory system's response to sound. Essentially, a signal passes via a pre-emphasis filter, then is divided into overlapping frames, and then each frame is subjected to a window function. Each frame is then subjected to a Fourier transform, or more precisely, a Fast Fourier Transform (FFT), which yields the energy spectrum. Finally, the filter banks are computed. To generate MFCCs, the filter banks undergo a Discrete Cosine Transform (DCT), with some of the resultant coefficients being retained and the remainder being discarded. Mean normalization is the last step applied [21]. Consult Figure 7 below for the diagram of this process.

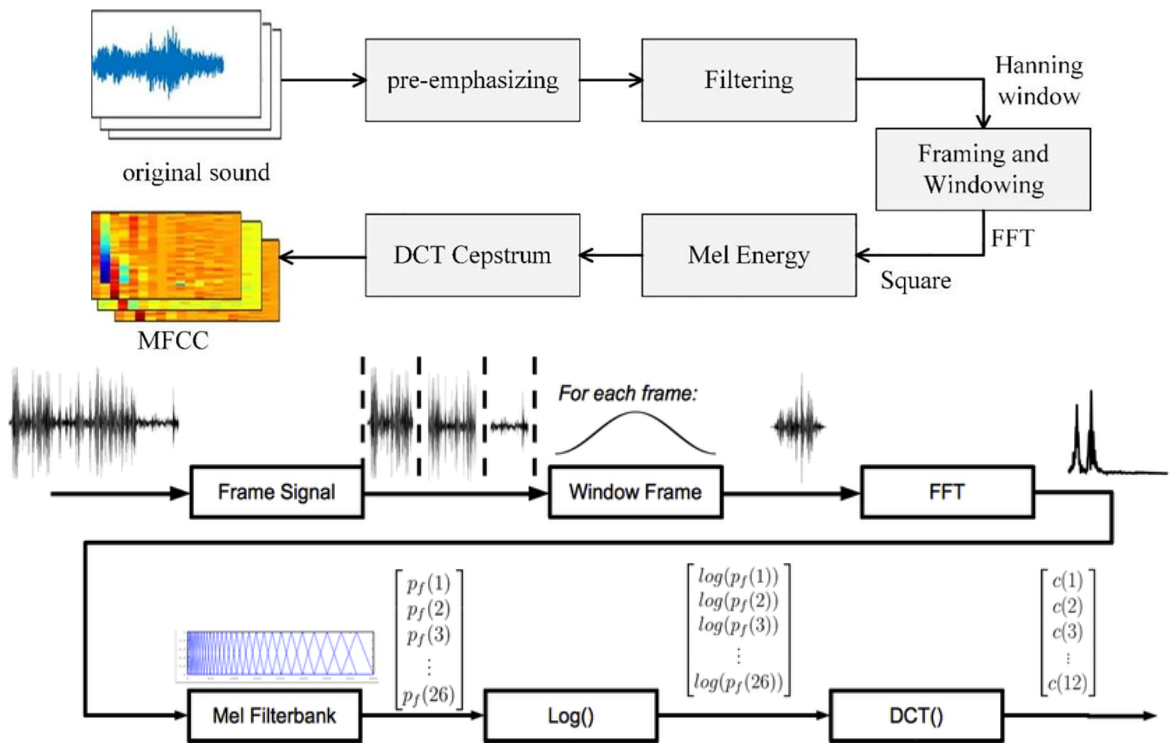

*Figure 7: Two MFCC calculation example diagrams* [22, 23].

MFCCs are fundamental in distinguishing instruments in various settings. For instance, a study by Nagawade and Ratnaparkhe in 2017 demonstrated the efficacy of MFCCs in identifying musical instruments from solo recordings. Their system utilized kNN classifiers and achieved high accuracy rates, highlighting MFCCs' capability to capture unique acoustic signatures of different instruments, such as the cello, piano, and trumpet [24]. Further research has explored combining MFCCs with other features for improved classification accuracy. For example, combining MFCCs with spectral features like sonograms has been shown to enhance the performance of classification systems using ML models like SVMs and kNN, as discussed in studies that tested these combinations on a variety of musical instruments and achieved notable accuracy improvements [25]. On this basis, the inclusion of this feature is a necessity for the classification. Consult Figure 8 below to see the visual representations of the MFCCs.

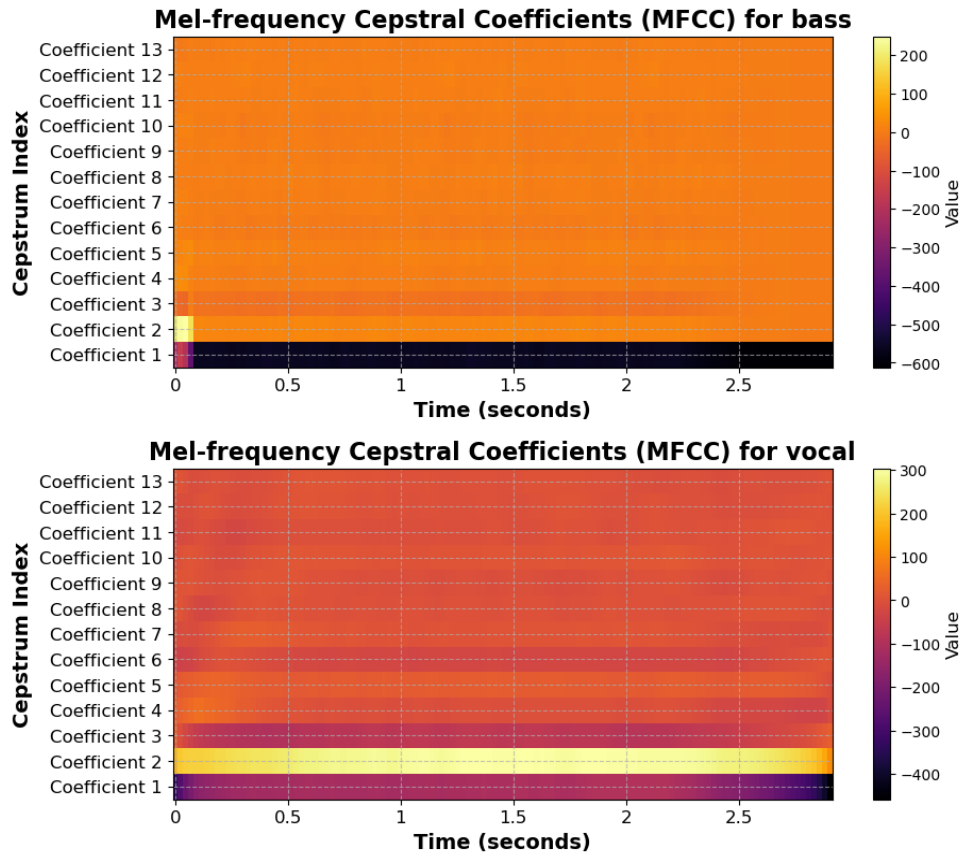

*Figure 8: MFCC representations of two samples of instruments.*

### 2.2.5 Spectral Contrast

Spectral contrast features are also significant in audio analysis because they provide a representation of the sound's spectral characteristics by highlighting the differences between peak and valley energies across different frequency bands. This method emphasizes the relative distribution of spectral energy, which can be crucially different between different types of musical instruments.

In their 2002 study, Dan-Ning Jiang et al. investigated the effectiveness of octave-based spectral contrast features in classifying different music types [26]. Their research focused on enhancing the management of digital music databases by providing a more robust method for music classification. The study introduced spectral contrast features that capture the variance between the most prominent peaks and the deepest valleys in the spectral energy distribution across various octave bands. This method contrasts with traditional approaches like MFCCs, which generally summarize the average spectral shape. Jiang et al. conducted experiments that demonstrated the superior capability of spectral contrast features in distinguishing between different music types. They showed that these features provided better discrimination among various musical styles compared to MFCC. The results highlighted the potential of spectral contrast features to enhance automatic music classification systems, making them more efficient in recognizing and categorizing diverse music genres [26]. This result is very likely to be applicable to a task such as instrument classification.

In 2010, Seo introduced Spectral Contrast MFCC (SCMFCC), a hybrid feature combining spectral contrast and traditional MFCC to improve music genre classification. This integration captures both the detailed spectral peaks and valleys and the averaged spectral envelope, offering a richer and more descriptive feature set. The study demonstrated that the SCMFCC excels in accessing genre-specific information, thus significantly enhancing the accuracy of music classification systems. By effectively identifying subtle differences between genres, SCMFCC proved superior to using MFCC alone, leading to a marked improvement in classification performance across various musical instrument styled genres [27]. Even though SCMFCCs were not implemented explicitly as a feature, the study still outlines the importance of spectral contrast in combination with MFCCs on audio classification tasks. Thus, this makes spectral contrast an invaluable tool in the domain of MIR, aiding in the effective classification of instruments. Consult Figure 9 below to see the visual representations of the spectral contrast of the samples.

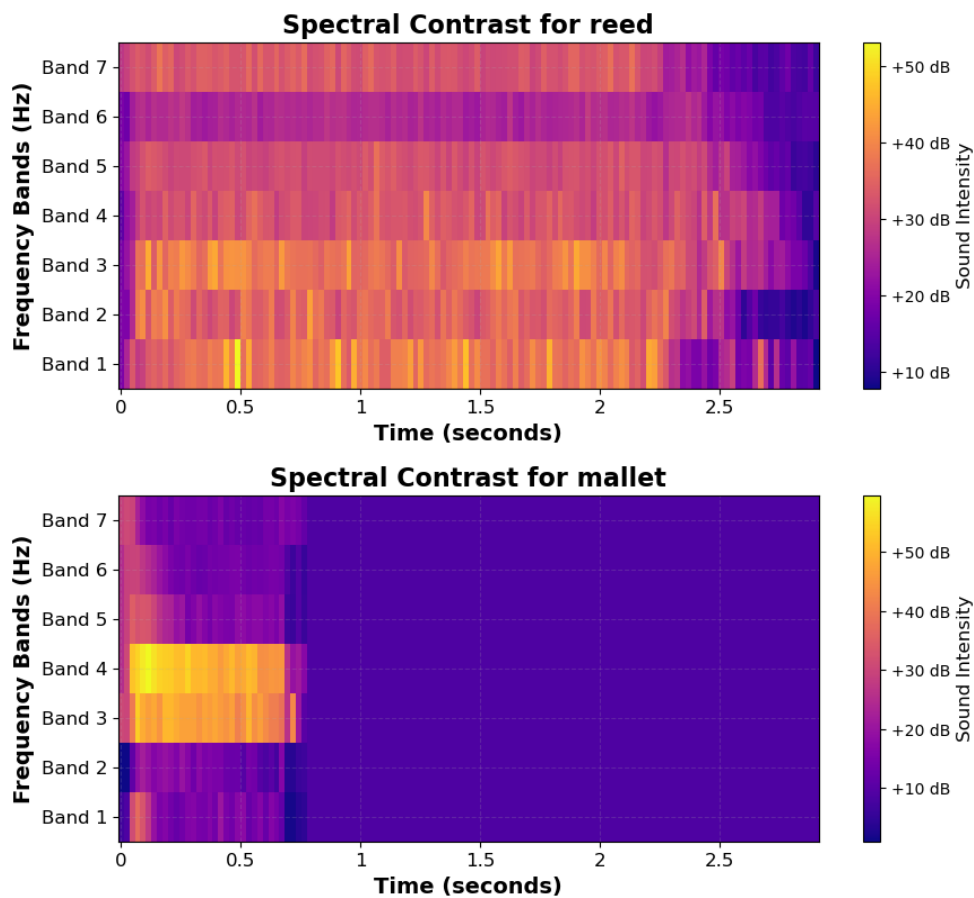

*Figure 9: Spectral contrast representations of two samples of instruments.*

Certainly, combining these five audio characteristics will provide strong results. Every single feature is essential because it captures unique aspects of audio signals and adds value to the analysis. Together, these elements create an adaptable and extensive toolbox that greatly enhances the capacity to identify and decipher even the most minute subtleties included in the audio data. This improved toolkit will hopefully make it possible to gain more accurate and thorough insights, leading to a deeper comprehension of the difficulties associated with audio signal processing.

## 2.3     Numerical Feature Data Wrangling

Having defined all the features that will be utilized to extract information from audio files, the next step is to structure our feature dataset effectively by exploring and analyzing the contents of the NSynth dataset for our task, which is available in JSON format. NSynth provides a rich array of audio samples along with metadata that includes various attributes such as acoustic characteristics, instrument type, and pitch. We need to examine this metadata to understand the data, which is crucial for the selection of samples. We will extract the names of the WAV file samples from the JSON file for efficient listing when extracting the features.

When working with audio files, especially in the context of extracting meaningful features for MIR, analysis and ML, the Python library Librosa offers a powerful set of tools. For our feature extraction process, we will utilize Librosa [28]. Librosa simplifies the process of feature extraction from audio signals by providing high-level functions specifically designed for audio and music analysis. For instance, to extract the MFCC, Librosa provides a function that computes these coefficients directly from the audio signal [28]. This process is similar for every other feature that we defined for the extraction process.

Read more about the Librosa library documentation here: https://librosa.org/doc/latest/index.html.

Instrument family classes are numbered as follows: 0 - Bass, 1 - Brass, 2 - Flute, 3 - Guitar, 4 - Keyboard, 5 - Mallet, 6 - Organ, 7 - Reed, 8 - String, 9 - Synth Lead, 10 - Vocal. Consult Figures 10, 11, 12 below showcasing the distribution of samples within the NSynth dataset.

_Note, this is for the raw NSynth dataset and not our adapted feature dataset._

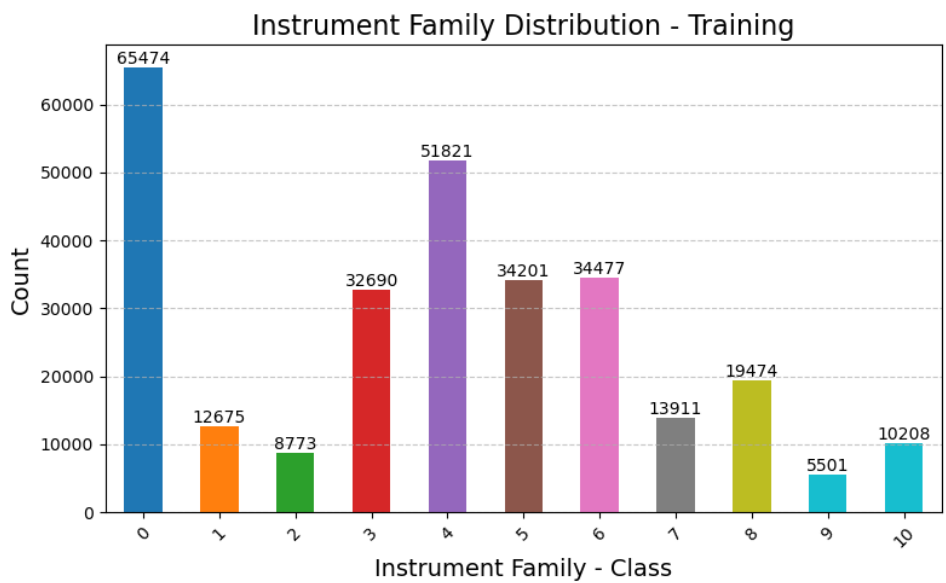

*Figure 10: Distribution of the number of samples in each instrument family class in the raw NSynth training dataset.*

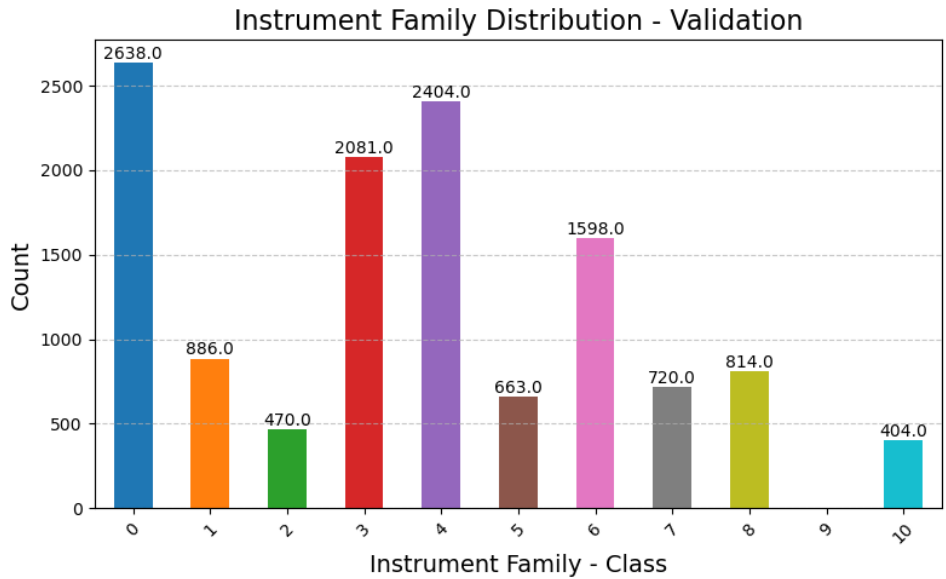

*Figure 11: Distribution of the number of samples in each instrument family class in the raw NSynth validation dataset.*

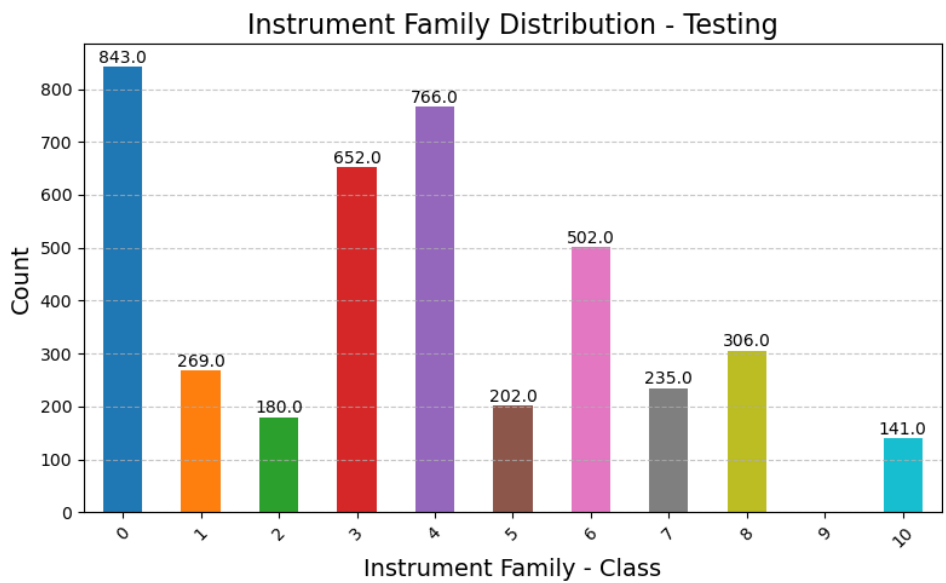

*Figure 12: Distribution of the number of samples in each instrument family class in the raw NSynth testing dataset.*

As we can see the bass class and the keyboard class in all datasets have the most samples. Flute and synth_lead have the lowest number of samples in all datasets. We see that synth_lead is missing samples in the validation and testing sets, as well as only containing a small amount of 5501 samples for training making it an obsolete inclusion for our feature dataset.

Conclusively, *synth_lead will not be included in the classification task* due to lack of data. After removing synth_lead the lowest number of samples belongs to the flute class. Following this change we are left with 10 instrument classes.

It is crucial to define the number of samples in each instrument class. This approach ensures that our model is trained on a balanced dataset, which is fundamental for achieving reliable and generalizable results. A balanced dataset helps prevent the model from being biased toward overrepresented classes and improves its ability to accurately classify less common instruments. In a study regarding imbalanced data, the authors explored several techniques for addressing class imbalance, including resampling the dataset to achieve uniformity across classes, and applying algorithmic adjustments that adapt the learning process to better handle imbalanced data. Their findings suggest that methods which create a more balanced class distribution tend to result in improved model accuracy and generalization capabilities, especially in scenarios where minority class instances are crucial for the predictive task [29]. Thus, our goal is to achieve a uniform distribution across all categories, thereby providing the classification algorithm with an equal opportunity to learn distinctive features from each instrument type.

To do this we will define datasets of varying equal sample sizes for each class so that we can also examine how the amount of data impacts the classification of musical instruments. Even though, it is well established that more data given to a model almost always gives better results.

The study by Banko and Brill is often cited for demonstrating the relationship between dataset size and ML model performance. The researchers explored how increasing the size of training datasets impacts the accuracy of NLP models and ML models. Their findings clearly indicated that performance improves consistently with the addition of more data, often overshadowing the gains from more sophisticated algorithms [30]. Furthermore, in a 2017 study, Sun et al. revisit the concept first explored by Banko and Brill, specifically in the context of DL. They systematically examine how increasing the size of datasets influences the performance of DL models across various tasks, including image recognition, NLP, and speech recognition. Their analysis shows that DL models continue to benefit from larger datasets, often achieving new state-of-the-art results as the amount of data increases [31].

We created 4 training feature datasets: 2000 samples per class, 3000 samples per class, 5000 samples per class and 8500 samples per class. The choice of a maximum of 8500 samples is due to the limit of the 8773 samples that we have for the flute class in the training set. For convenience, we named the datasets NuFDIC – Numerical Feature Dataset for Instrument Classification. They are comprised of the same 5 feature elements we defined earlier and have a uniform distribution of the samples for each class. However, these 4 datasets differ in general sample size. To ensure consistency of samples we checked for the distribution of the sources of the instruments. Source classes are numbered as follows: 0 - Acoustic, 1 - Electronic, 2 - Synthetic. Moreover, we also checked for quality reassurance within all instrument samples by checking the sample rate consistency within the datasets.

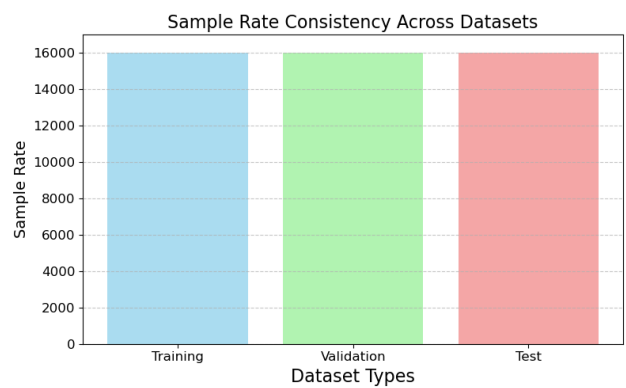

*Figure 13: Sample rate of all samples in the raw NSynth datasets.*

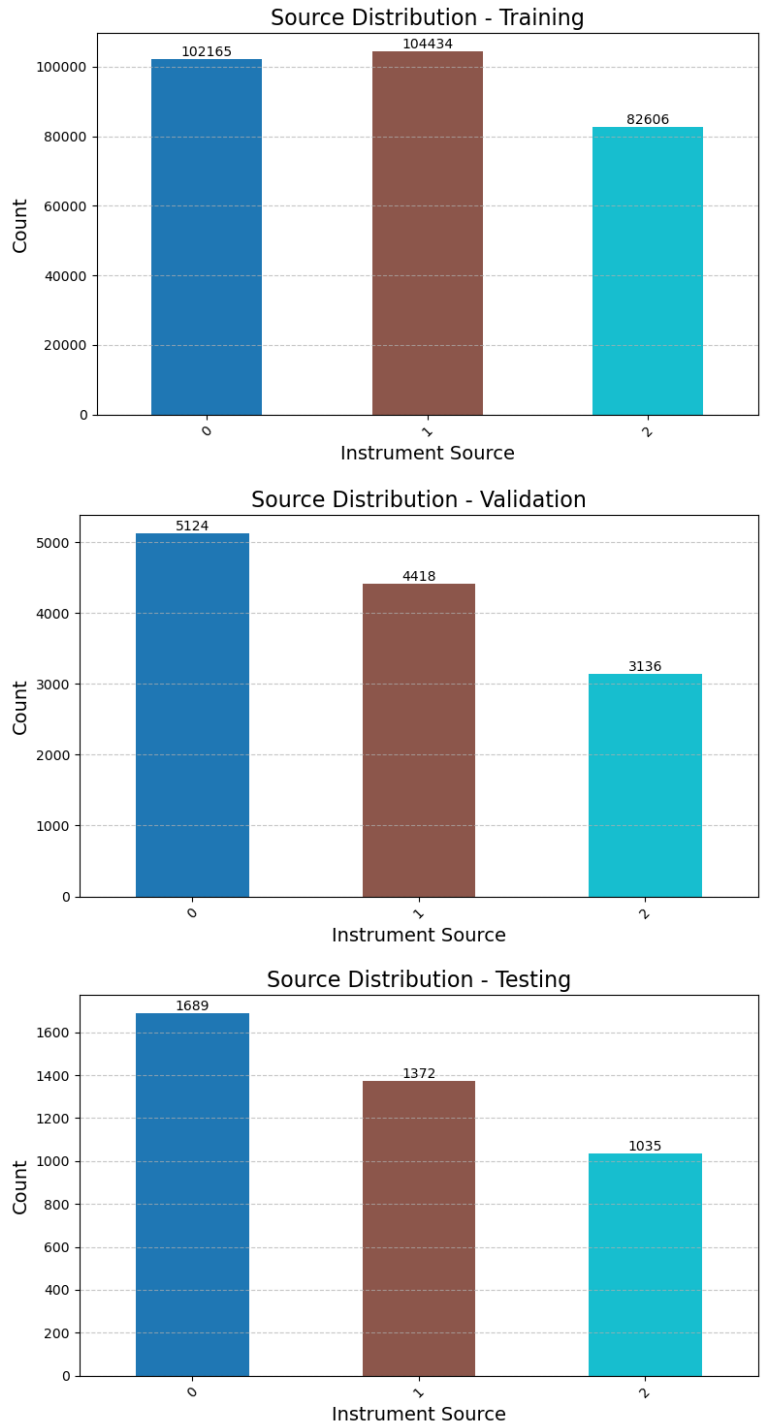

*Figure 14: Distribution of the number of samples in each instrument source class in the raw NSynth datasets.*

Note, we do not need to display and focus on the pitch or the velocity of the samples as it is not something we consider for the classification of instrument families. Although after analysis the velocity followed a well-defined uniform distribution for all datasets. The pitches were also distributed well resembling a Gaussian Distribution for all datasets.

## 2.4    Spectrogram Image Data Creation and Analysis

We opted to also create more simple data for evaluation, instead of the complex feature dataset we created earlier, for a different classification approach we will use images. We created image data for evaluating models on a different data type to get a new perspective. We again created four distinct datasets, each comprising different numbers of samples: 2000, 3000, 5000, and 8500. For convenience, we named the datasets SIDIC – Spectrogram Image Dataset for Instrument Classification.

The choice of using spectrograms instead of Mel spectrograms was preferred.

Mel spectrograms compress the frequency information by scaling it logarithmically, which, while efficient for many applications like speech recognition, inherently leads to some loss of the original signal's information. This compression is beneficial for reducing the data size and simplifying the feature space for ML models, making computations faster and less resource intensive. However, this compression becomes a double-edged sword when the task requires high fidelity in audio reconstruction. In contrast to Mel spectrograms, standard spectrograms maintain a linear representation of frequency and time, preserving more detailed information about the sound signal. Therefore, when audio quality and precise reproduction of the original sound are paramount, such as in scientific analysis where every nuance of the sound might carry important information, then the use of standard spectrograms is often preferable. The research highlights a critical trade-off between efficiency and accuracy, pointing to the need for careful selection of spectrogram type based on the specific requirements and goals of the audio processing task [32].

Thus, spectrograms were selected for evaluation due to their broader historical usage in a lot of studies. As we mentioned, traditional spectrograms display a frequency spectrum over time at linear scales, providing a different representation in image format rather than Mel spectrograms, having a detailed and unaltered visual representation of sound, capturing nuances that are crucial for accurate sound analysis. Ideally, this reasoning will lead to better results.

In their 2017 study, Costa et al. explore the application of CNNs in music and sound analysis, focusing on the use of spectrograms as inputs for tasks such as music genre classification, instrument recognition, and sound quality assessment. Spectrograms provide a rich source of information captured in visual patterns that CNNs can effectively interpret. By training CNNs on these visual representations, the networks are able to autonomously identify and learn significant features from complex audio signals. This process eliminates the need for manual feature extraction, traditionally a labor-intensive aspect of sound analysis. The CNNs' ability to extract nuanced features from spectrograms not only streamlines the analytical process but also significantly improves the accuracy and efficiency of classifying and assessing different types of audio content. This technique has broad implications, demonstrating high effectiveness in diverse music and sound scenarios, making it a valuable tool for advancing audio analysis technologies [33].

Despite the availability of other visual audio representations (some of which we visualized earlier in section 2.2) including MFCCs, chroma images, Constant-Q Transform (CQT) images, Zero Crossing Rate (ZCR) representations, spectral contrast images and Tonnetz representations, were not chosen due to a lack of comprehensive comparative evidence demonstrating their superiority or equivalency to regular spectrograms or Mel spectrograms for the intended tasks.

Consult Figure 15 to see differences between the two spectrogram types and to see examples of the data. For the actual training data, image quality of 100 dpi was selected, resulting in 775 x 308 pixels.

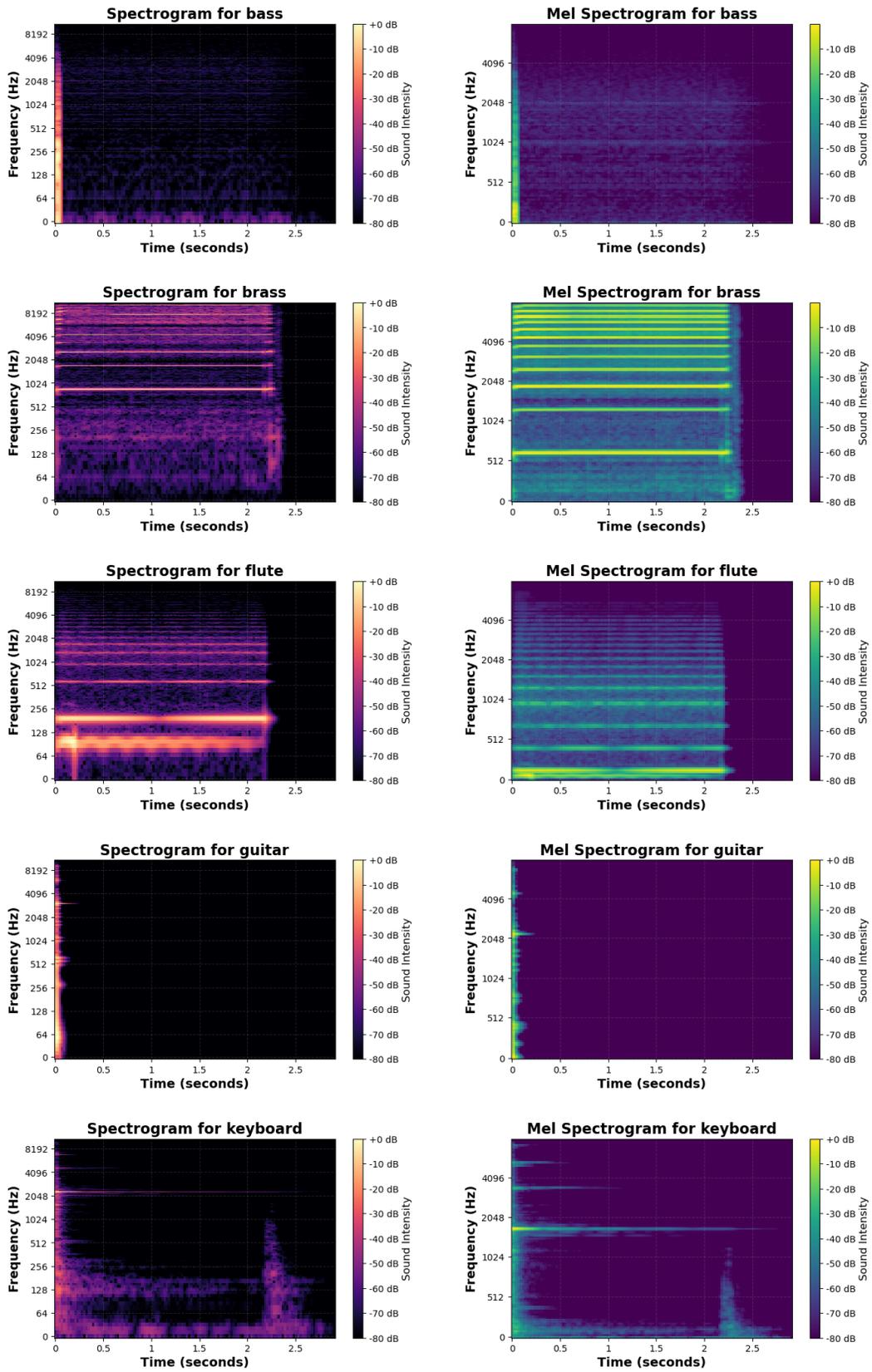

*Figure 15: Examples of spectrogram and Mel spectrogram images of the audio samples.*

# 3 Advanced Machine and Deep Learning Modeling

Advancements in ML has transformed the field of audio analysis, providing advanced solutions for signal processing, synthesis, classification, and more. For the task of instrument classification, it is more suitable to use supervised ML rather than unsupervised ML to get better more reliable results. In addition, NSynth itself is a supervised dataset with given metadata and labels to each sample.

NuFDIC as well as SIDIC which we created prior from NSynth are supervised datasets and categorized in 10 different classes. The aim is to compare the best supervised ML models for our task and to aim for the best measured results, while contrasting differences.

Note, even though the models and methods we used for comparison are labeled in different sections with different names, they are all supervised ML methods, including DL which is part of ML.

## 3.1 Supervised Machine Learning Methods

In this section we explore more "traditional" standalone ML methods to classify instruments. Our NuFDIC dataset will be utilized with the methods in this section.

One of the more basic classifiers, Naïve Bayes (NB) is usually used as a classifier for various studies. The classifiers work by employing probabilistic models that integrate the detailed features to make more accurate predictions about the music's attributes and classifications. As demonstrated in the work by Fu, Lu, Ting, and Zhang in 2010, this method has shown decent performance in identifying and retrieving music based on genre [34]. Although in their study they classified music genres using adapted and more advanced NB methods. Our task is an auditory classification task similar to theirs, thus using a simple NB as a guideline comparison point was a viable option.

We will also utilize an advanced form of a Bagging classifier - Random Forest (RF). In the study of environmental audio classification, RF classifiers have demonstrated significant advantages over traditional mathematical prediction methods. Zhang and Lv in 2015 explored the efficacy of RF in handling environmental audio data, which is often complex and noisy, making it challenging to classify using standard methods. The strength of RF in this application lies in their ability to manage high-dimensional data and their robustness against overfitting [35]. RF operate by constructing multiple decision trees during the training phase. Each tree is built on a random subset of the data features and training examples, and it makes its classification based on those subsets. This randomness helps in reducing the variance of the model, thus improving the generalization of the classifier over unseen data. Furthermore, one of the most valuable aspects of the RF algorithm is its ability to evaluate the importance of different features for classification. The key to the success of RF in audio classification is their method of feature selection and utilization. Because we are working with many different features this will be extremely useful.

In the context of RF, hyperparameter optimization is crucial because it significantly impacts model accuracy and generalizability. RFs involve several hyperparameters like the number of trees, maximum depth of the trees, minimum samples split, and the number of features to consider when looking for the best split. Selecting the optimal values for these parameters can be challenging and computationally expensive.

Random search (RS) is a practical and efficient method for hyperparameter optimization in ML models, including RF classifiers. Unlike grid search, which systematically explores a predefined range of hyperparameter values, random search selects hyperparameters randomly from a defined search space and evaluates their performance. This approach can lead to better results within a fraction of the computation time required by more exhaustive methods. Due to the costs of grid search, we opted to use RS in our RF classifiers – RFRS. Note, we will also evaluate a default RF model.

In a 2007 study, Zhu, Ming, and Huang explored the effectiveness of SVM for audio genre classification within a content-based video retrieval framework. Their approach utilized SVM to classify audio signals into six distinct classes: pure speech, music, silence, environmental sound, speech with music, and speech with environmental sound. The rationale for using SVM in this context stems from its capability in handling high-dimensional data and its capability to model complex nonlinear decision boundaries efficiently. SVM's strength in delivering higher classification accuracy compared to other methods such as decision trees, kNN, and neural networks was particularly highlighted [36]. In our context, the study demonstrated that SVM are viable when classifying audio into relevant categories. This makes SVM a suitable choice for tasks requiring audio analysis within multimedia application.

## 3.2    Boosting Algorithms Implementation

Boosting algorithms are a family of ML algorithms designed to improve the accuracy and performance of predictive models. The central principle of boosting is to combine weak learners to form a strong learner. A weak learner is defined as a classifier that performs slightly better than random guessing. Boosting algorithms iteratively learn weak classifiers with respect to a distribution and add them to a final strong classifier. After each round, it increases the emphasis on misclassified instances so that subsequent classifiers focus more on difficult cases. Our NuFDIC dataset will be utilized with the methods in this section.

AdaBoost, short for Adaptive Boosting, has proven highly effective in various classification tasks, including music classification from audio waveforms. In a study by Bergstra et al. in 2006, AdaBoost was utilized to select and combine audio features extracted and aggregated from segmented audio for musical genre and artist prediction. The key to their method's success in winning in the MIREX 2005 international contests on MIR was AdaBoost's ability to iteratively focus on difficult examples by adjusting the weights of classifiers and features in each round of learning. This approach is particularly suitable for audio classification because it can handle varied and complex data structures in audio files efficiently. The adaptability of AdaBoost to different types of features and its mechanisms against overfitting make it an ideal choice for tasks where distinguishing between subtle differences in audio can be crucial [37]. Its proven track record in handling complex classification problems, provides a solid choice for our task.

Currently, there seems to be limited direct research on using Gradient Boosting (GB) specifically for audio classification. GB is a powerful ML technique known for its prediction accuracy and ability to handle heterogeneous features, but its application in audio processing isn't as extensively documented in the provided results from recent academic studies.

Although, comparisons to AdaBoost have been shown in several studies and competitions. In a 2019 study by Bahad and Saxena, the comparative effectiveness of GB and AdaBoost was rigorously tested using a two-class predictive model on a widely used diabetes dataset. This research highlighted the superior performance of GB over AdaBoost in terms of prediction accuracy. The study utilized ensemble machine learning techniques, where both algorithms were applied to classify instances as either presenting with or without diabetes based on various risk factors. GB's ability to outperform AdaBoost was attributed to its optimization framework, which iteratively minimizes a specified loss function, effectively handling both bias and variance in the dataset. This capability allows GB to improve prediction accuracy significantly, particularly in scenarios where understanding intricate patterns in data is crucial for classification accuracy [38]. Proving to be a newer, "better" and more applicable approach to a broader range of problems we also evaluated GB in our task.

EXtreme Gradient Boosting (XGB) is an implementation of GB that has been optimized for speed and superior performance. XGB includes several enhancements (Regularization, Tree Pruning, Handling Missing Values, System Optimization) that differentiate it from conventional GB.

In a 2022 study by Liu, Yin, Zhu, and Cui, a novel approach to musical instrument classification was developed, utilizing the strengths of XGB in conjunction with feature fusion. This method involved extracting a diverse set of audio features from various channels, which were then ingeniously fused to create a comprehensive feature set that provided a detailed representation of audio signals. The fusion of features such as timbre, pitch, and rhythm, among others, allowed the XGB algorithm to effectively distinguish between different musical instruments, significantly improving the classification accuracy. By training the XGB model with these complex, fused datasets, the researchers achieved an impressive classification accuracy of 97.65% on the Medley-solos-DB dataset, surpassing the performance of traditional classification models [39]. Conclusively, XGB will prove to be a great addition for evaluation in instrument classification.

## 3.3    ANN Deep Learning Modeling

Artificial Neural Networks (ANNs) are computational models inspired by the human brain's structure and function. They are designed to recognize patterns and solve complex problems by mimicking the way neurons interact in biological brains.

DL is a subset of ML that employs deep neural networks, which are ANNs with multiple hidden layers (at least 1 hidden layer to be deep). In recent years, this method of solving problems has gained a lot of attention with the globalization of AI. Our NuFDIC dataset will be utilized with the models in this section.

In a 2021 study by Mahanta, Khilji, and Pakray, a DL model was specifically designed to tackle the challenge of musical instrument recognition by utilizing MFCCs as key input features. The ANN was trained on a comprehensive dataset comprising recordings from the London Philharmonic Orchestra, which provided a rich variety of instrumental sounds. This diversity was critical as it exposed the model to a broad spectrum of acoustic nuances, thereby enhancing its ability to generalize across different instruments in varied orchestral contexts. The model was deemed successful with 97% validation accuracy [40]. The findings from this study emphasize the potential of ANNs in enhancing the classification of complex instrumental audio data. Hence, we will include ANN models for our study task.

For evaluation we created 3 different ANN models with different attributes. We labeled them Simple, Dropout and Complex according to their features. Consult Figure 16 below to examine the ANN models in detail with their respective attributes.

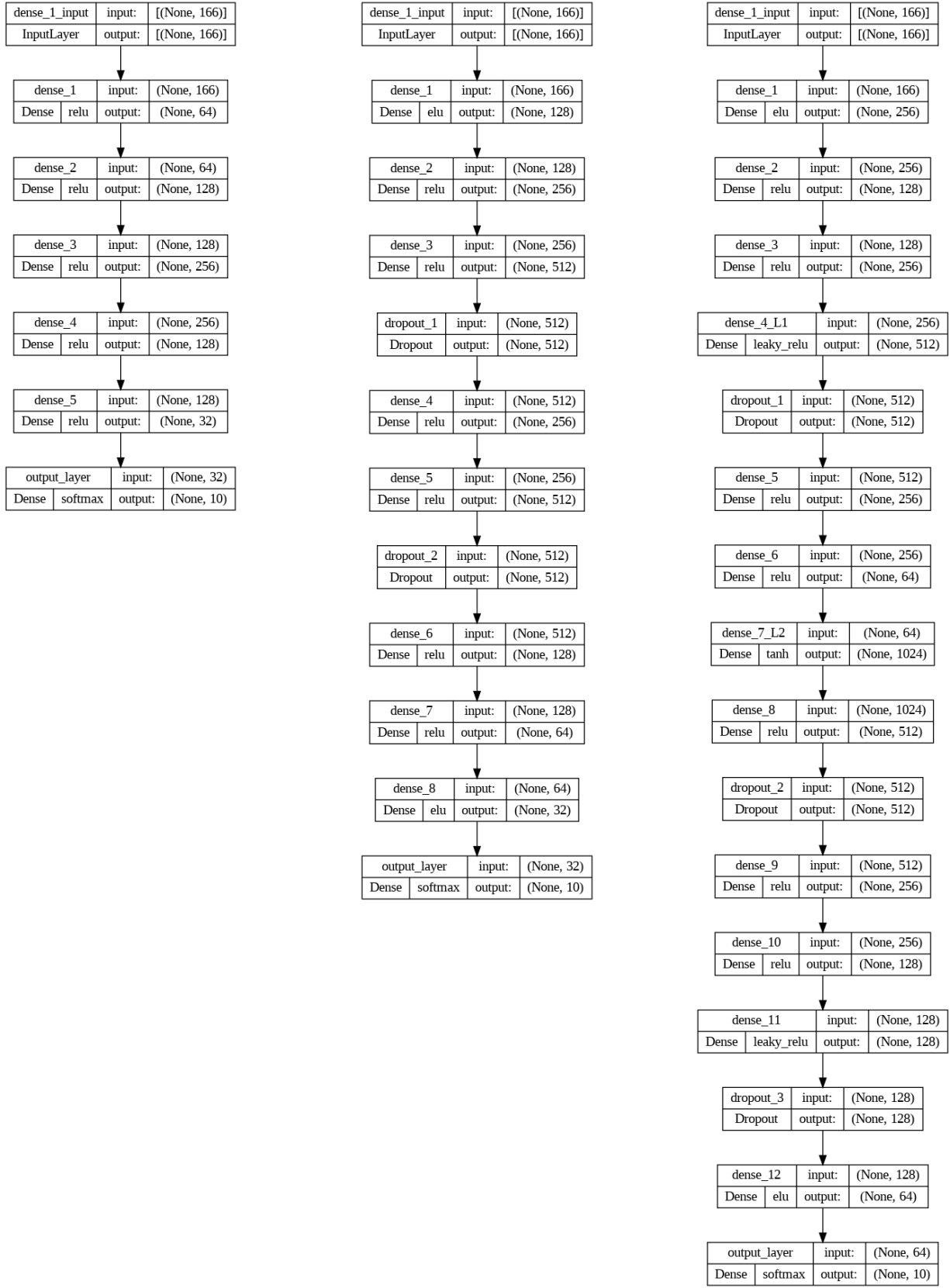

*Figure 16: Overview of our Simple, Dropout and Complex DL ANN Models displayed in that order.*

The simple ANN model serves as a foundational approach to understanding the behavior of neural networks under minimal complexity. This model consists of a few dense layers with a moderate number of neurons. Its primary purpose is to establish a baseline performance, offering insights into how well basic neural structures can capture and model the relationships. By employing fewer parameters, the simple model reduces computational demands as well.

To enhance the performance of ANNs and combat overfitting, a dropout model was introduced. Dropout is a regularization technique where randomly selected neurons are ignored during training, meaning their contribution to the activation of downstream neurons is temporally removed on the forward pass and any weight updates are not applied to the neuron on the backward pass. This method helps in preventing neurons from co-adapting too much, thereby forcing the network to learn more features that are useful in conjunction with many different random subsets of the other neurons. By dropping out neurons, the model learns to generalize better, improving its performance on unseen data.

The complex model involves deeper architectures with additional layers and more neurons, incorporating functionalities like batch normalization and advanced activation functions. This model is aimed at capturing more abstract patterns and relationships in the data, which may be missed by simpler models. While it has a higher capacity for learning, it also poses a greater risk of overfitting. To mitigate these risks, our complex model employs regularization techniques, including L1/L2 regularization and dropout, as well as techniques to ensure stable convergence, like careful initialization of weights and learning rate adjustments.

Both the dropout and complex model were introduced to combat overfitting, underfitting and to improve generalization. We will evaluate if the complex model proves to be too advanced for our data with results.

We did not incorporate recurrency in our models or create RNN models due to limitations on our data as well as the burden of time and cost limitations. Regarding their implementation in MIR, fellow colleagues/professors from Utrecht University have done excellent work regarding this topic. The paper "Musical Instrument Classification using Democratic Liquid State Machines" by Jornt R. de Gruijl and Marco A. Wiering from Utrecht University, published in 2006, introduces the Democratic Liquid State Machine (DLSM). This innovative neural network architecture employs an ensemble of Liquid State Machines (LSMs) for enhanced performance in time-series data analysis, particularly in classifying musical instruments by timbre. LSMs integrate a recurrent spiking network, a type of RNN known as the "liquid," effectively capturing dynamic states over time. This architecture is particularly effective in tasks involving complex temporal patterns, such as sound recognition. The study demonstrates that while individual LSMs achieved a classification accuracy of 96% for instruments like bass guitar and flute, the DLSM approach increased accuracy to 99%, showcasing substantial improvements in performance and learning capabilities over traditional neural network models [41]. Decisively, if we were to structure the data differently and adapt it for the use of RNNs and their variations, promising results would not be a surprise.

## 3.4    CNN Deep Learning Modeling

Historically, the use of CNNs in music classification began to gain significant traction in the early 2010s, as researchers discovered their potential in image processing could be translated into audio analysis by treating spectrograms as images. The use of spectrograms in audio analysis for classification tasks has been around since the early 2000s. This approach allowed for the automatic extraction and learning of features from complex audio data, a task that traditional ML methods handled less efficiently.

A very recent study conducted by Yuqing Su in 2023 evaluates various ML models on their effectiveness in classifying musical instruments using monophonic sound samples. This comprehensive analysis includes models like kNN, SVM, Gaussian Mixture Modeling (GMM), ANNs, CNNs, and RNNs. The study employs methods we previously detailed in our paper, in various sections above, except for GMM. These methods have consistently shown strong performance in MIR, as demonstrated in several studies we referenced. Among these, CNN emerged as the most accurate, achieving an impressive accuracy of 96.82%, while SVM was identified as the fastest, processing data in just under a minute. The performance of the CNN was particularly notable, especially when considering the classification of instruments that produce similar timbres, such as trumpets and saxophones. The study employed 1D-CNN with modifications to the layer structure and feature set, experimenting with both 1D and 2D layers and varying the number of MFCC features. These factors combined make the CNN approach stand out in the study as particularly effective for the classification of musical instruments, showcasing its potential to be the best choice, especially in applications where high accuracy is the main criteria, such as ours [42].

For evaluation, we developed a custom CNN IC model tailored for our task on the SIDIC dataset. This model processes input images which are preprocessed and normalized. The architecture comprises multiple convolutional layers with 32 and 128 filters, using ELU and ReLU activations to introduce non-linearity and mitigate vanishing gradients. Regularization techniques such as L2 and L1 are employed in convolutional and dense layers, respectively, to prevent overfitting by penalizing complexity. Spatial dimension reduction is achieved through successive max-pooling layers. The network features several dense layers with dropout mechanisms to prevent overfitting. A variety of node sizes are used to capture features effectively (64 to 1024). Consult Figure 17 below to see an overview of our CNN IC model.

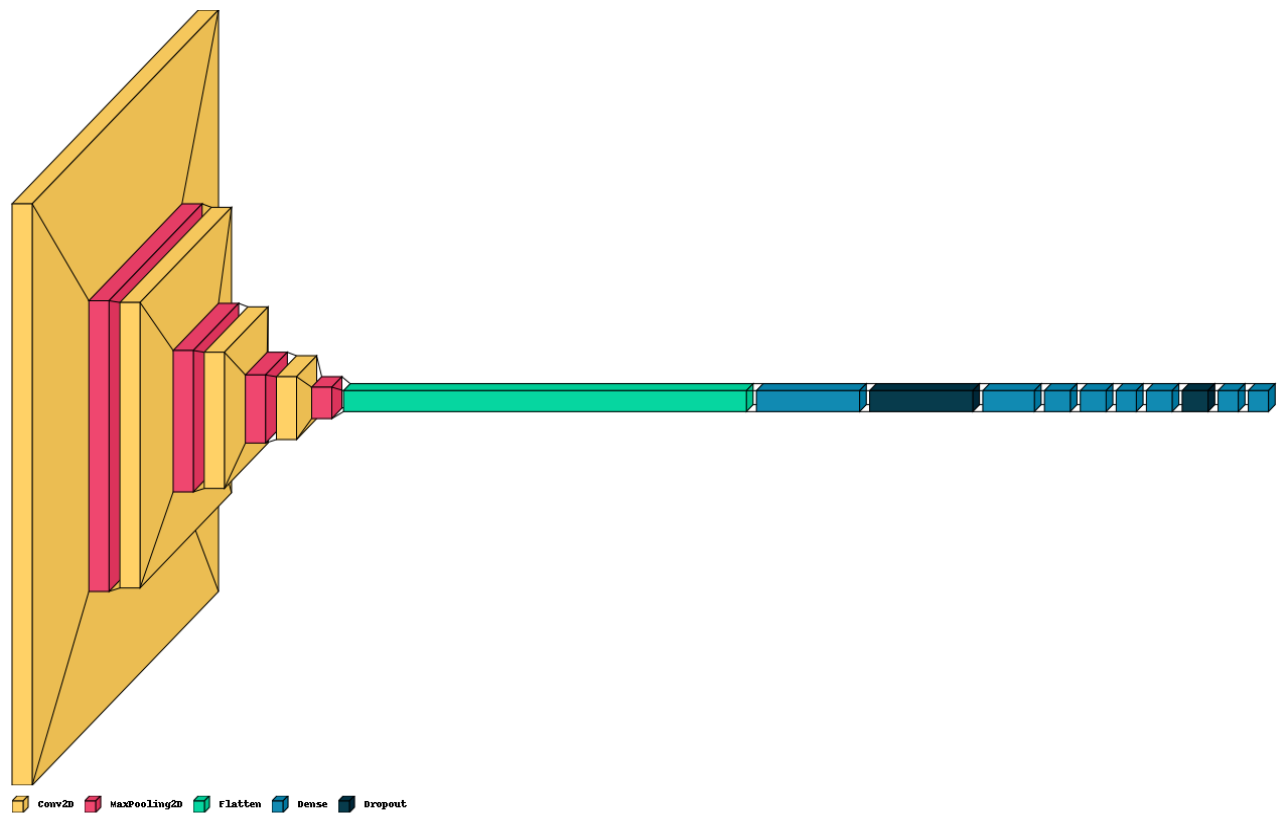

*Figure 17: Overview of our CNN IC model for instrument classification.*

# 4 Results and Model Accuracies

In this section we will discuss the results for each method. We will also mention characteristics such as tested and used hyperparameters, batch sizes, epochs and model configurations. Optimizers will also be discussed for models that use them, such as in the case of ANNs and CNNs. We will utilize 2 methods to assess the results we get from the models. We will evaluate performance based on a corresponding confusion matrix and an accuracy rate.

A confusion matrix is a square matrix of size N x N, where N represents the total number of distinct target classes in a classification problem. It is used to show the counts of true positives (TP), true negatives (TN), false positives (FP), and false negatives (FN) for each class. Thus, it helps in understanding both the effectiveness of the model and the specific types of errors it might be making by comparing predicted labels against the actual ones [43]. The confusion matrix is a useful and comprehensive presentation of the classifier performance. It is commonly used in the evaluation of multi-class, single-label classification models, where each data instance can belong to just one class at any given point in time [44].

Now that the notions of TP, TN, FP, FN have been defined we can define the common ML measure of accuracy. Accuracy is a metric that measures how often a model correctly predicts the correct outcome, hence how accurate it is in its job. We will measure accuracy on both the test and validation set. Accuracy is given by the formula:

$$Accuracy = \frac{TP + TN}{N}$$

In the ML community, due in part to the realization that simple classification accuracy is often a poor standalone metric for measuring performance, we introduced the confusion matrix representation as well [45]. We concluded it is needed for better evaluation of our results. Even though, other metrics exist we decided to utilize these as their use is well-established and optimal for our case. Consult Figure 18 below to see the logic of a confusion matrix and other metrics that can be used for performance evaluation.

True class

|  | p | n |
|--|--|--|
| Y | True Positives | False Positives |
| N | False Negatives | True Negatives |

Hypothesized class

Column totals: P    N

$\text{fp rate} = \frac{FP}{N}$    $\text{tp rate} = \frac{TP}{P}$

$\text{precision} = \frac{TP}{TP+FP}$    $\text{recall} = \frac{TP}{P}$

$\text{accuracy} = \frac{TP+TN}{P+N}$

$\text{F-measure} = \frac{2}{1/\text{precision}+1/\text{recall}}$

*Figure 18: Confusion matrix and metrics commonly used in ML, including accuracy which we will use [45].*

## 4.1    Supervised Machine Learning Methods – Results Evaluation

We present our results for the NB, RF, RFRS, SVM classifiers in separate tables. We will present only the best models' confusion matrix for each method.

For the NB classifier only a standalone default model was trained. Consult Table 3 below for accuracy results.

*Table 3: Accuracy results of the NB classifier.*

| Model NB | NuFDIC Dataset | | | |
|---|---|---|---|---|
| | 2000 Samples | 3000 Samples | 5000 Samples | 8500 Samples |
| **Accuracy** | 12.92 | 14.06 | 12.52 | 18.5 |

We can clearly see our NB classifier performs very poorly with all data sample sizes, reaching a peak accuracy of 18.5 on the largest sample size, which is a bit better than random guessing. We can see where the model makes errors in the confusion matrix below.

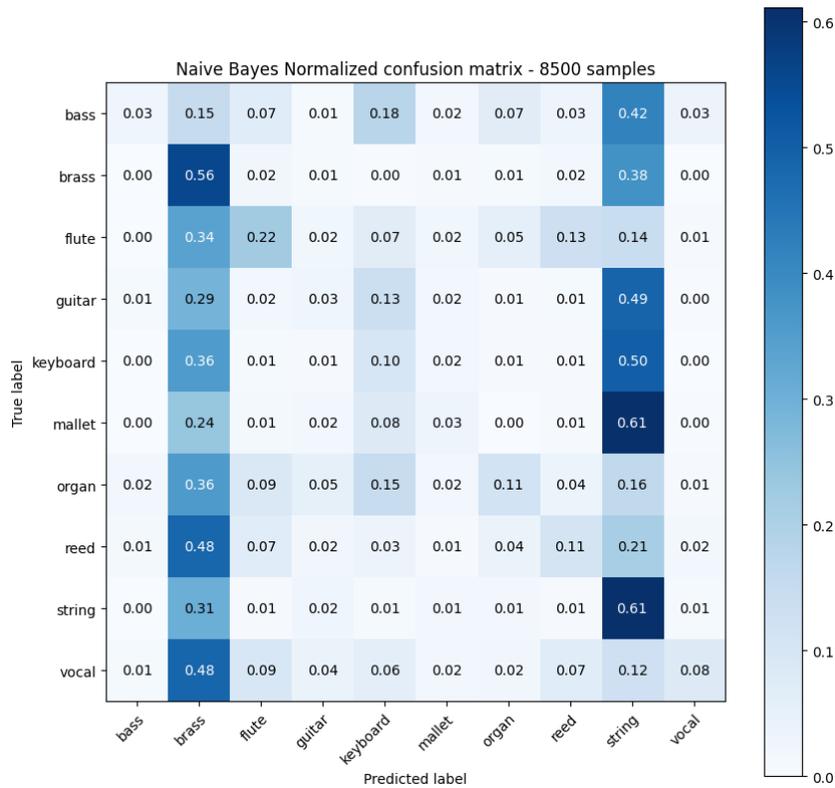

*Figure 19: Confusion matrix of the NB model trained on 8500 samples.*

For our RF model, we implemented it using unoptimized and default hyperparameters with a total of 100 trees. Consult Table 4 below for accuracy results.

*Table 4: Accuracy results of the RF classifier.*

| Model RF | NuFDIC Dataset | | | |
|---|---|---|---|---|
| | 2000 Samples | 3000 Samples | 5000 Samples | 8500 Samples |
| **Accuracy** | $55.26 \pm 0.46$ | $58.6 \pm 0.22$ | $60.28 \pm 0.41$ | $91.85 \pm 0.24$ |

We can see a massive increase in performance when compared to the NB model, with very satisfactory results when trained on 8500 samples. The confusion matrix is given below.

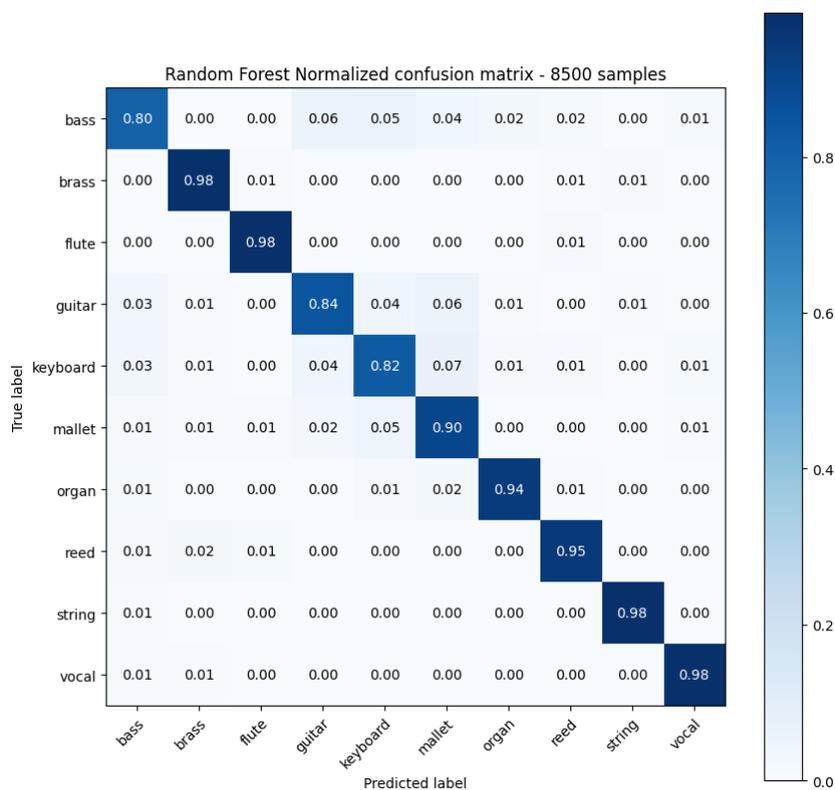

*Figure 20: Confusion matrix of the RF model trained on 8500 samples.*

We can see the model performs very efficiently on a large dataset, as it rarely misclassifies most samples. The only problems it encounters is the "3 x 3 bad performance square" with center at the keyboard class on the diagonal in the matrix. This makes some sense since in fact higher pitched guitar, mallet and piano sounds do in fact sound very similar. We will focus on this in our results to establish if it is a trend or just a specific case for the RF model.

We implemented the RS hyperparameter set with various options, multiple times to get a better average for expected results and to test more options, since RF are very susceptible to hyperparameter settings. Some of the hyperparameters we tested include bootstrap options, depth of trees, maximum features per tree for classification, number of leaf nodes, and criterion for purity of splits. After testing, the best model used the highest number of trees, totaling 512, with a maximum depth of 64. See Table 5 below for results.

*Table 5: Accuracy results of the RFRS classifier.*

| Model RFRS | NuFDIC Dataset | | | |
|---|---|---|---|---|
| | 2000 Samples | 3000 Samples | 5000 Samples | 8500 Samples |
| **Accuracy** | $56.37 \pm 0.58$ | $58.41 \pm 0.47$ | $61.5 \pm 0.5$ | $93.59 \pm 0.69$ |

We see similar performance to the RF model, although the accuracy when training the model on 8500 samples gives an accuracy of 93.59, which is the best yet. Thus, RFRS yielded a significant boost in results compared to RF, which did not have optimized hyperparameters. The confusion matrix is given below.

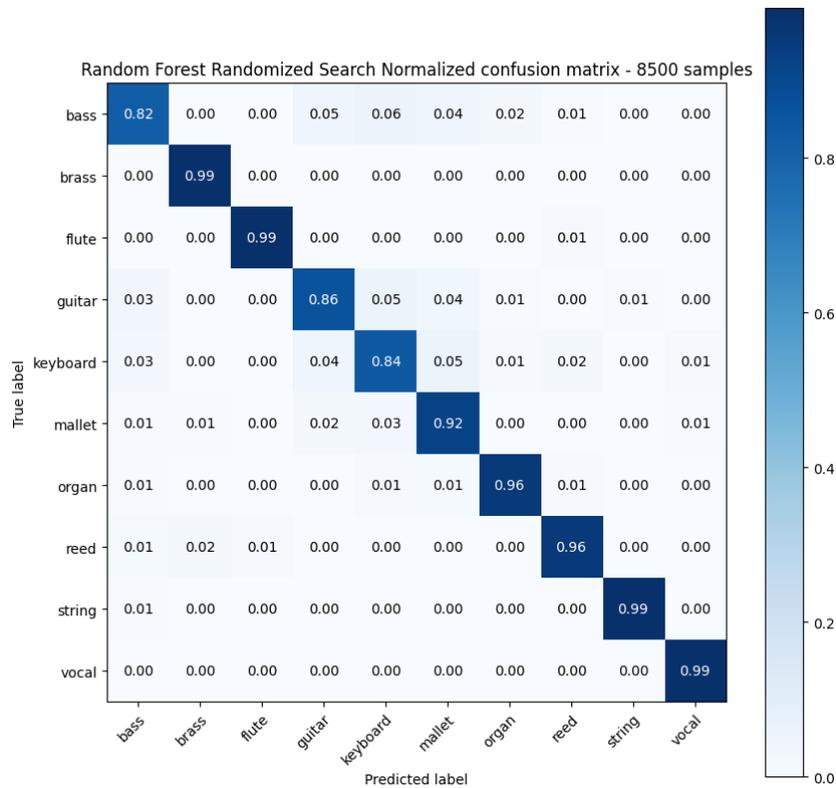

*Figure 21: Confusion matrix of the RFRS model trained on 8500 samples.*

The model reached almost perfect classification on some instruments with rates up to 0.99. Note that the "3 x 3 bad performance square" is still present at the keyboard class on the diagonal.

In SVMs, the *c* hyperparameter is a regularization hyperparameter that controls the trade-off between achieving a low error on the training data and minimizing the model complexity for better generalization to new data. It plays a role in controlling the decision boundary in the algorithm. A lower value of *c* in SVMs can often be better for generalizing to new data, although it can lead to underfitting, resulting in higher errors on both training and validation datasets. On the other hand, a very high *c* value minimizes the number of misclassifications on the training data, which might lead to overfitting. Ideally, the SVM model should have an optimal low *c* value with the least error on training and validation sets [36]. We decided to set *c* to a commonly used value of 10. We used the Radial Basis Function kernel. See Table 6 below for results.

*Table 6: Accuracy results of the SVM classifier.*

| Model SVM | NuFDIC Dataset | | | |
|---|---|---|---|---|
| | 2000 Samples | 3000 Samples | 5000 Samples | 8500 Samples |
| **Accuracy** | 39.6 | 39.62 | 40.38 | 75.62 |

The SVM model did not achieve better results than RFRS. It capped out at an accuracy of 75.62 with the biggest sample size. The confusion matrix is given below.

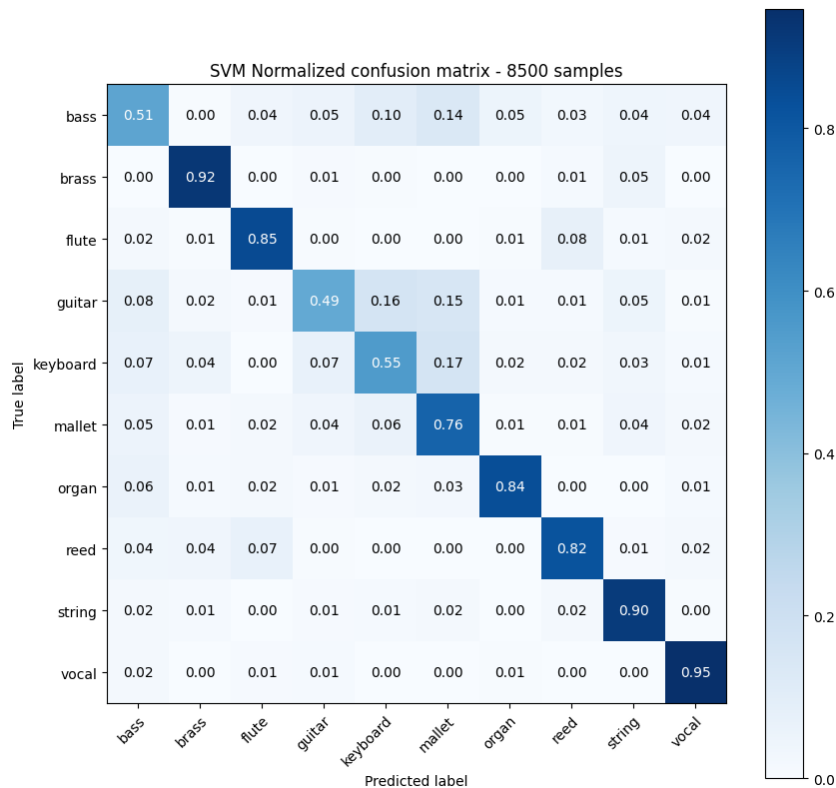

*Figure 22: Confusion matrix of the SVM model trained on 8500 samples.*

We can see resemblance of the trend we defined earlier. This model also shows more errors with the bass.

## 4.2    Boosting Algorithms Implementation – Results Evaluation

See result Table 7 for the AdaBoost classifier below.

*Table 7: Accuracy results of the AdaBoost classifier.*

| Model AdaBoost | NuFDIC Dataset | | | |
|---|---|---|---|---|
| | 2000 Samples | 3000 Samples | 5000 Samples | 8500 Samples |
| **Accuracy** | 42.32 ± 0.07 | 42.59 ± 0.36 | 28.95 ± 0.96 | 48.03 ± 0.04 |

AdaBoost surprisingly showed unsatisfactory results for all sample sets. Especially for the 5000 samples per class set. Due to reasons unknown, the accuracy suffered significantly over several runs. Currently we are not able to formulate a hypothesis on what the reason for this is, because there is a jump in accuracy again on the 8500 sample set. The 8500 sample set hit the highest accuracy of only 48.03. The confusion matrix may provide some insight.

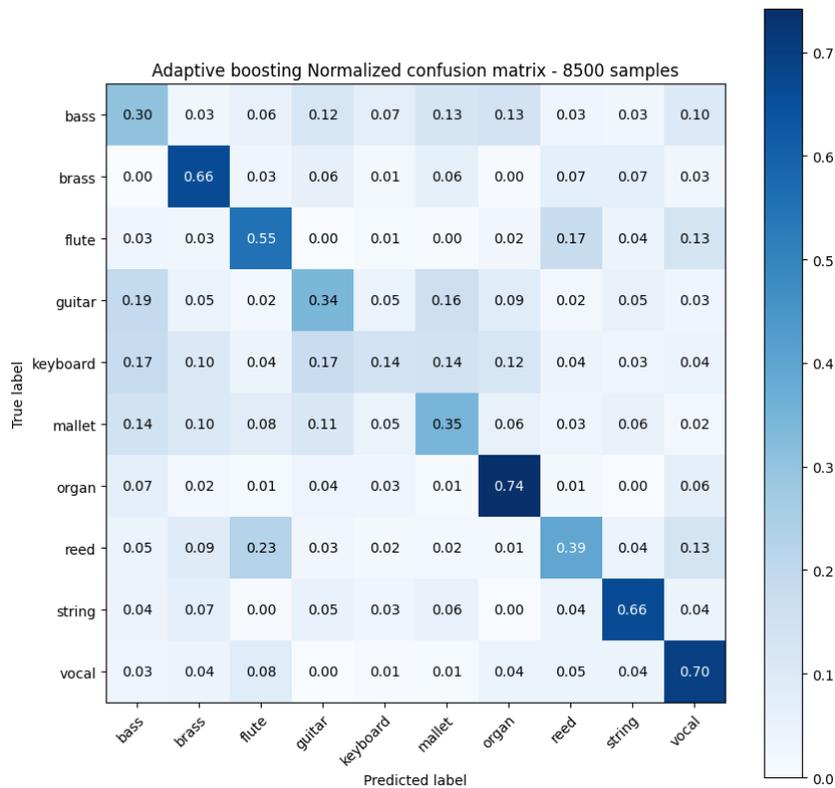

*Figure 23: Confusion matrix of the AdaBoost model trained on 8500 samples.*

Predictions are weak overall. Note, we can still see the keyboard trend present in other methods than RF.

See result Table 8 for the GB classifier below.

*Table 8: Accuracy results of the GB classifier.*

| Model GB | NuFDIC Dataset | | | |
|---|---|---|---|---|
| | 2000 Samples | 3000 Samples | 5000 Samples | 8500 Samples |
| **Accuracy** | $42.17 \pm 0.35$ | $42.21 \pm 2.05$ | $40.29 \pm 0.17$ | $49.83 \pm 0.28$ |

Still these results disappoint in comparison to some of the methods we presented. Although, no real abnormalities are observed, as the accuracies seem to follow the basic trend of increasing as we reach 8500 samples. Although the increase between 2000, 3000, 5000 and 8500 samples is not substantial enough for the introduced computational time when training on more samples. See the confusion matrix below.

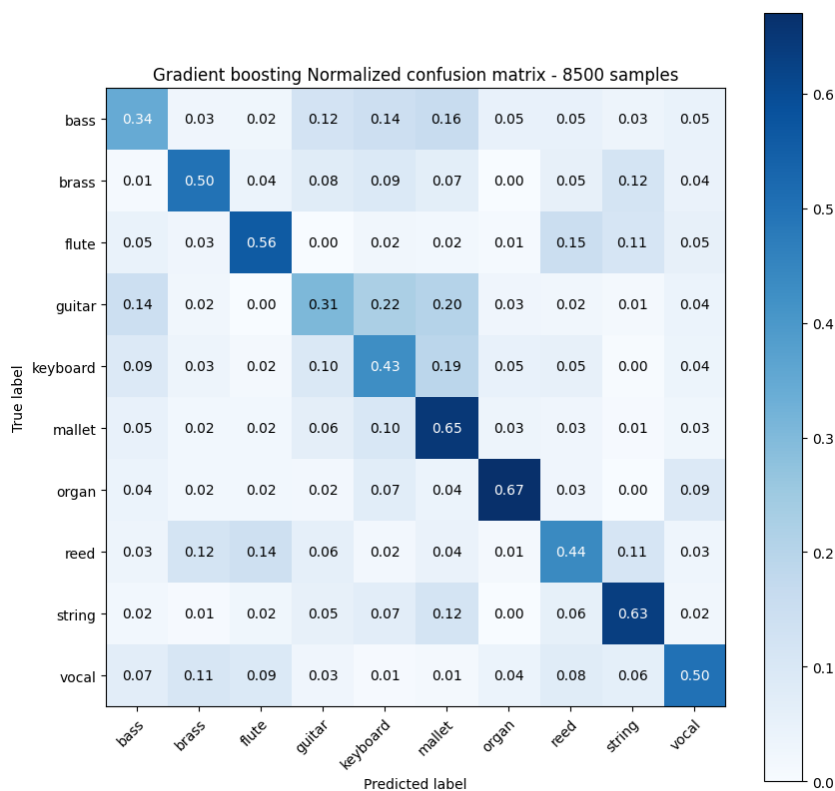

*Figure 24: Confusion matrix of the GB model trained on 8500 samples.*

The model poorly classifies bass and guitar samples. As with the models before it has the trend of the "3 x 3 bad performance square" which indicates that keyboard, mallet and guitar confusion is also present within the matrix. Due to this, the guitar class gets often classified as a keyboard or mallet by error.

Finally, the XGB classifier, often considered the successor of the GB classifier. See result Table 9 below.

*Table 9: Accuracy results of the XGB classifier.*

| Model XGB | NuFDIC Dataset | | | |
|---|---|---|---|---|
| | 2000 Samples | 3000 Samples | 5000 Samples | 8500 Samples |
| **Accuracy** | $56.69 \pm 0.18$ | $58.05 \pm 0.13$ | $58.45 \pm 0.83$ | $89.29 \pm 1.59$ |

The results over all sample sizes show improvements over both AdaBoost and GB. Even the model trained on 2000 samples hit better accuracy than the best models of AdaBoost and GB, which were trained on 8500 samples. The best accuracy is again reached using 8500 samples, totaling 89.29. Various combinations of hyperparameters were tested. Drop rate was also included for regularization.

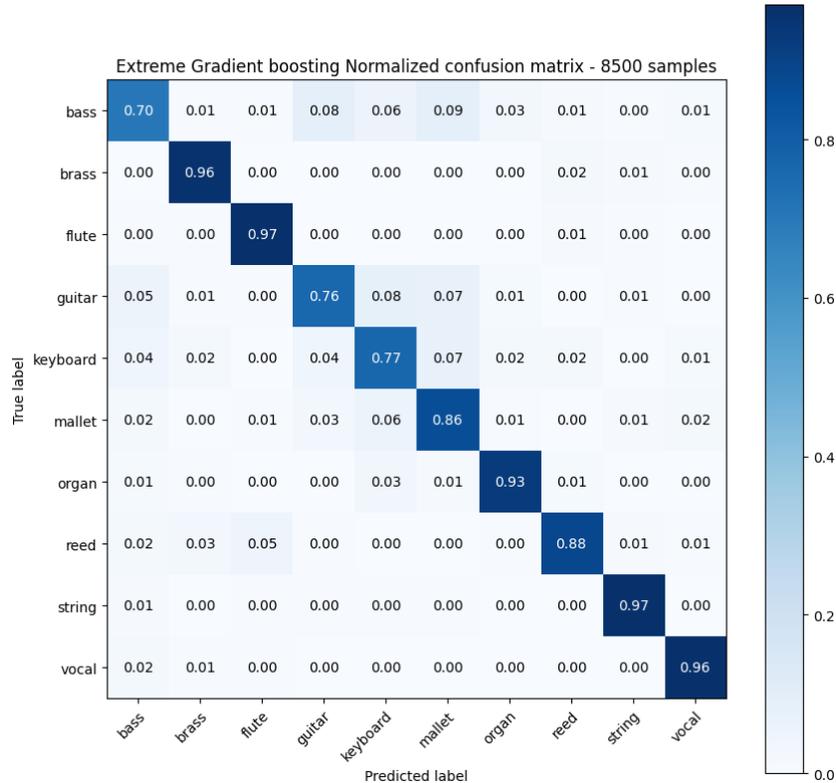

*Figure 25: Confusion matrix of the XGB model trained on 8500 samples.*

This model performed best from all the other boosting models. It can predict 5 classes with accuracy above 0.93. It again struggles with bass predictions, like some of the previous models. Also, we can see the same trend with errors surrounding the keyboard class.

## 4.3    ANN Deep Learning Models – Results Evaluation

When we trained our ANN models, we trained them on varying batch sizes of 16, 32, 64, 128 with various epochs to prevent overfitting. Training epochs ranged from 20 to 1000 during testing. Most models seemed to converge to a stable validation accuracy at around 100 to 300 epochs. As for model configuration, various implementations were tested. We implemented a learning rate scheduler which adjusts the learning rate during training, which can help improve the model's performance and convergence speed. We also implemented early stopping to monitor if the models' performance on the validation set starts to degrade. In our models, we utilized categorical cross-entropy as the loss function, which is particularly suited for multi-class classification problems where class membership is mutually exclusive, such as in our case. For optimizer choice we decided that Adam is the best choice since its introduction in 2015 [46]. Although the effectiveness of optimization algorithms can vary depending on the specifics of a given problem, Adam often outperforms other optimizers like SGD, AdaGrad, and RMSProp in DL scenarios due to several key advantages. Adam combines the best properties of AdaGrad and RMSProp [46]. The use of all other optimizers we mentioned is still common, although we did not explicitly test them for this task. We will also display the loss and accuracy of the best performing models. See the SimpleANN results in Table 10 below.

*Table 10: Accuracy results of the SimpleANN classifier.*

| Model SimpleANN | NuFDIC Dataset | | | |
|---|---|---|---|---|
| | 2000 Samples | 3000 Samples | 5000 Samples | 8500 Samples |
| **Accuracy** | 40.56 ± 0.65 | 40.6 ± 0.56 | 41.22 ± 0.25 | 87.68 ± 0.94 |

Similar to the previously discussed methods we can see a sudden jump in accuracy when the model is trained on 8500 samples. When navigating the loss and accuracy landscapes we see that the model converges to a peak validation accuracy at around 80 -100 epochs, starting to bounce around a certain value afterwards. As for the validation loss, we can see it rise linearly after 40 epochs. When the validation loss rises while the accuracy remains stagnant, it often indicates that the model is struggling. Thus, our model beings capturing noise and irrelevant patterns along with the actual patterns. It performs exceptionally well on training based on the training loss and accuracy. The confusion matrix shows good results also. See figures 26, 27 bellow.

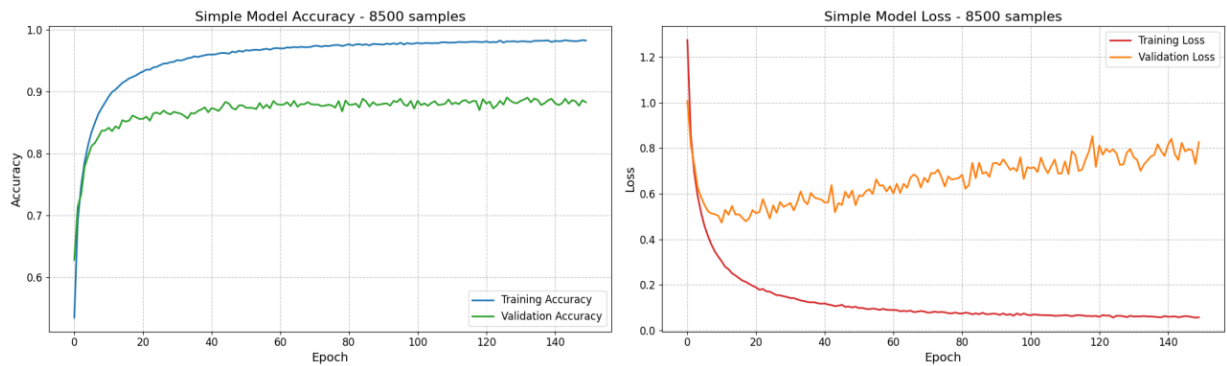

*Figure 26: Accuracy and Loss over epochs while training the SimpleANN on 8500 samples.*

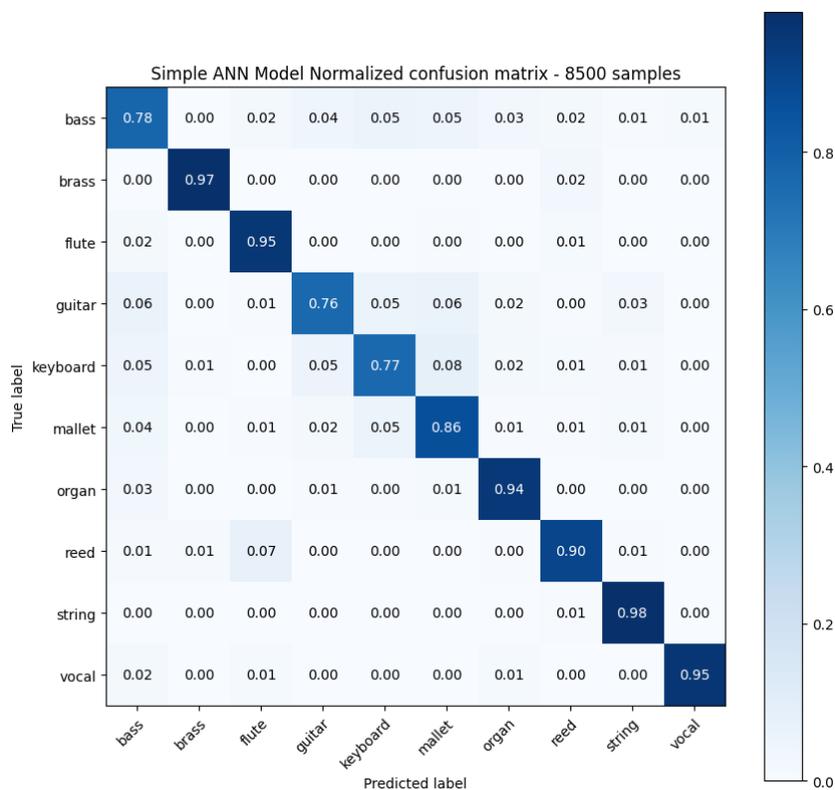

*Figure 27: Confusion matrix of the SmipleANN model trained on 8500 samples.*

As far as our ANN models, peak accuracy was achieved when training our DropoutANN model on 8500 samples. Yet, there is a sudden jump between 5000 and 8500 samples, whilst for 2000, 3000, 5000 samples the accuracy remains within the same range without sudden changes. See result Table 11 below.

*Table 11: Accuracy results of the DropoutANN classifier.*

| Model DropoutANN | NuFDIC Dataset | | | |
|---|---|---|---|---|
| | 2000 Samples | 3000 Samples | 5000 Samples | 8500 Samples |
| **Accuracy** | $40.98 \pm 0.05$ | $40.35 \pm 0.5$ | $39.4 \pm 0.82$ | $91.24 \pm 0.75$ |

When navigating the loss and accuracy landscapes of this model, both accuracy was higher and loss was lower when compared to the SimpleANN model. Here, the validation accuracy converged at around 200-250 epochs. The validation loss seems to hit a minimum at around 100 epochs, afterwards it starts linearly rising in a similar fashion to the SimpleANN model but in a much slower fashion due to the dropout layers.

In the confusion matrix we can see the model hitting near perfect accuracy for vocal, string, brass and organ classification. While still showing the trend surrounding the keyboard class that we see in all methods we previously described. See Figures 28. 29 below.

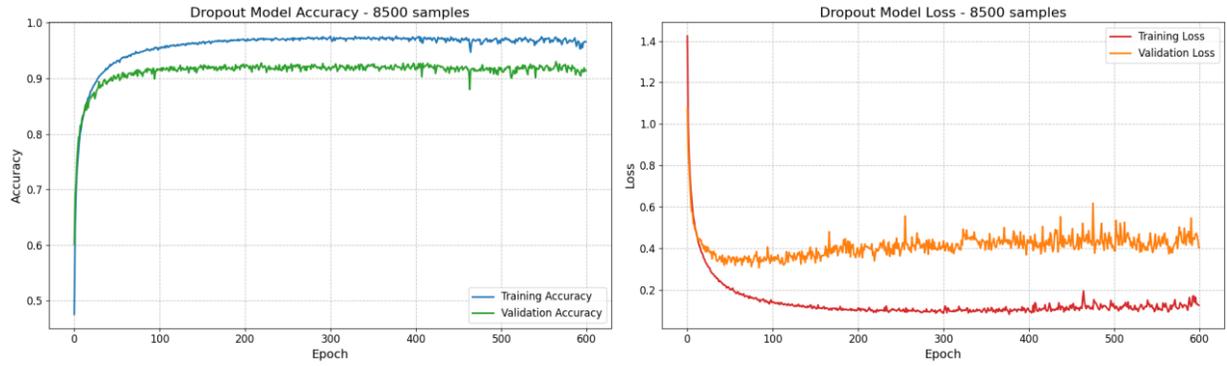

*Figure 28: Accuracy and Loss over epochs while training the DropoutANN on 8500 samples.*

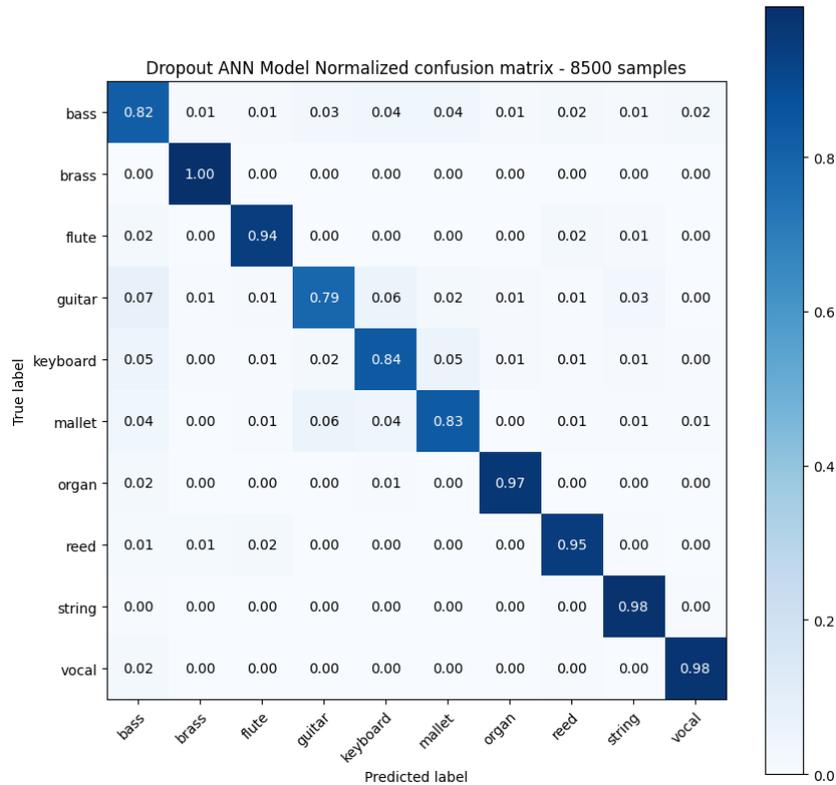

*Figure 29: Confusion matrix of the DropoutANN model trained on 8500 samples.*

The ComplexANN proved to be not as efficient as predicted in capturing the proper weights for our data. It does outperform the SimpleANN on 8500 samples but struggles to compete with the DropoutANN. Furthermore, it has lower accuracy when trained on the smaller sample amounts due to its complexity. Fewer layer implementations with perhaps more node depth would have performed better. Complex models also often come with a larger number of hyperparameters that need to be carefully tuned. Suboptimal hyperparameter settings can exacerbate issues, which might have been at fault. Regarding loss and accuracy, it is difficult to discuss convergence because the model does seem to slowly improve over all epochs. It does so linearly, but very slowly. Perhaps, running the model for more than a 1000 epochs would have led to a better model. This was not possible, due to cost and time constraints. See result Table 12 below.

*Table 12: Accuracy results of the ComplexANN classifier.*

| Model ComplexANN | NuFDIC Dataset | | | |
|---|---|---|---|---|
| | 2000 Samples | 3000 Samples | 5000 Samples | 8500 Samples |
| **Accuracy** | 36.95 ± 1.84 | 38.47 ± 0.54 | 36.87 ± 1.99 | 88.5 ± 1.68 |

Training loss and accuracy did deliver promising results. The keyboard square trend is still present in the confusion matrix, although 0.9 ± 0.9 prediction rate was hit on 6 classes. See Figures 30, 31 below.

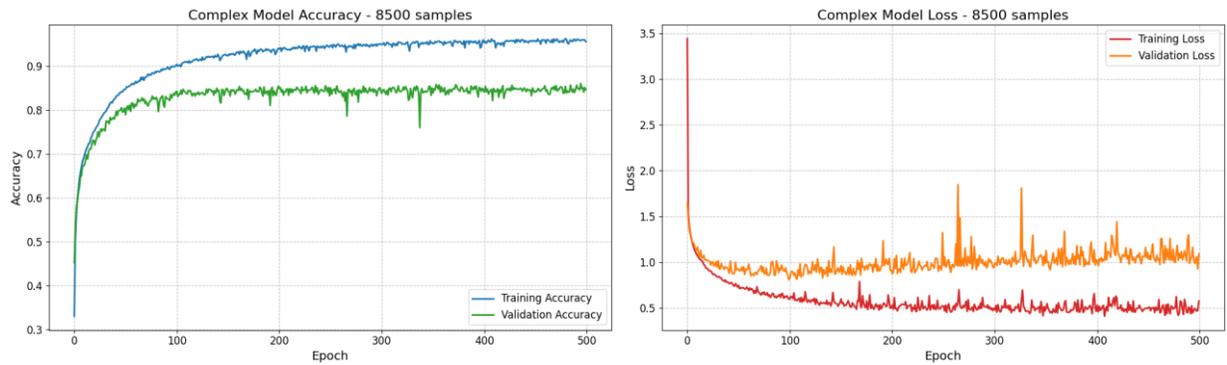

*Figure 30: Accuracy and Loss over epochs while training the ComplexANN on 8500 samples.*

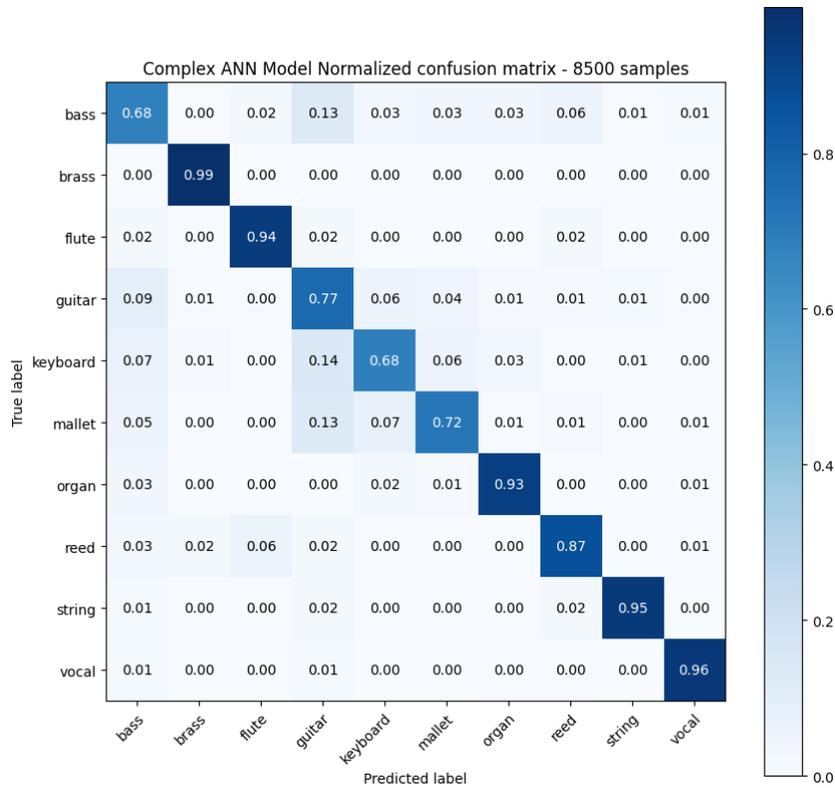

*Figure 31: Confusion matrix of the ComplexANN model trained on 8500 samples.*

## 4.4 CNN Deep Learning Models – Results Evaluation

In the development of our CNN IC model, we employed multiple training methods to optimize performance. We systematically experimented with varying batch sizes, specifically 64, 128, 256, and 512, to evaluate their impact on the model's learning dynamics and convergence rates. Training was conducted over a range of 20 to 100 epochs, allowing us to closely observe and determine the epoch at which convergence was typically achieved. Additionally, we explored a variety of filter sizes within the convolutional layers and experimented with different depths of dense nodes to refine feature extraction and decision processes. As the case is the same as for the ANN models, optimizer choice remained as Adam and categorical cross-entropy was chosen as the loss function.

After training, this model outperformed all other models for every sample size, respectively. Both validation and test accuracies were exceedingly high, with a peak of 98.72. This only further strengthens and proves that the use of image data for MIR is a very competitive and viable option. See result Table 13 below.

*Table 13: Accuracy results of the CNN IC classifier.*

| Model CNN IC | SIDIC Dataset | | | |
|---|---|---|---|---|
| | 2000 Samples | 3000 Samples | 5000 Samples | 8500 Samples |
| **Accuracy** | 88.75 ± 0.4 | 91.2 ± 0.25 | 95.88 ± 0.22 | 98.72 ± 0.17 |

Regarding convergence, early stopping seemed to evaluate that the model converged at around 77 epochs when trained on 8500 samples. Although, after manual examination, we can certainty state that the model could have been trained up to 100 epochs without the risk of overfitting. Thus, setting the patience sensitivity low for our early stopping function was at fault here. Unfortunately, we could have had the CNN IC model perform even better on 8500 samples. Training this model more than a very limited number of times was not possible though, due to time and cost constraints. For example, training only this specific version of the model took more than 30 minutes per epoch, resulting in a total of 40 hours.

Clearly, having the best performance in this case comes at a hefty price. Both time and computational resources were utilized heavily. See loss and accuracy in Figure 32 below.

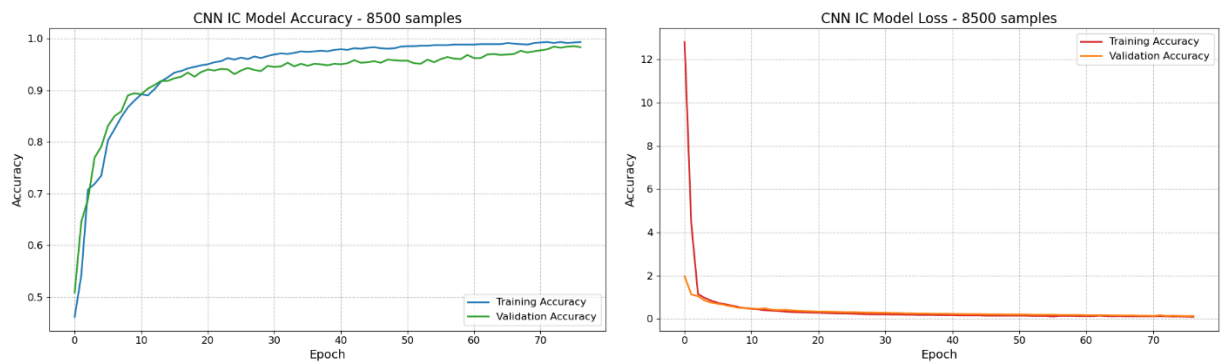

*Figure 32: Accuracy and Loss over epochs while training the CNN IC Model on 8500 image samples.*

The confusion matrix reveals that our model gets near perfect predictions on 5 classes. For the first time after the NB classifier (which was too weak to handle any trends), we do not see the trend of the "3 x 3 bad performance square" that we defined in section 4.1. Although, the most struggle the model has, is indeed with the mallet, keyboard and guitar classes. Note, the trend is also not present in the confusion matrix of the model trained on 5000 samples (we did not present this figure, as we deemed it was not necessary). Thus, the cutoff point for accurately distinguishing between mallet, guitar and keyboard sounds using the CNN IC model seems to be around 5000 image samples. See confusion matrix below.

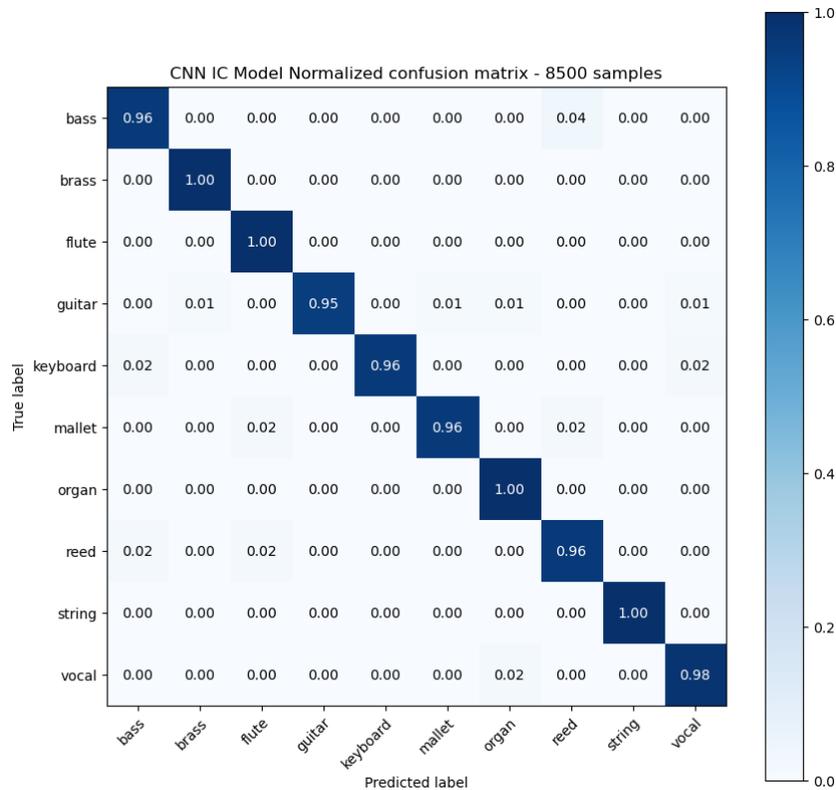

*Figure 33: Confusion matrix of the CNN IC model trained on 8500 image samples.*

# 5    Results Summary and Analysis

## 5.1    Overview of Classification Accuracies

We present an overview of all previously discussed results, which are comprehensively compiled in Table 14. This table provides the used classifier model, as shown in the leftmost bolded column. Additionally, the table includes details about the dataset utilized for each model and the specific sample sizes employed during training. Importantly, all numerical values listed within the table represent the classification accuracy rates of the respective models. For a detailed examination of these results and their implications, please refer to the information provided below.

Table 14: Overview of accuracy results of all respective models.

| Model | NuFDIC Dataset | | | |
|---|---|---|---|---|
| | 2000 Samples | 3000 Samples | 5000 Samples | 8500 Samples |
| NB | 12.92 | 14.06 | 12.52 | 18.5 |
| RF | 55.26 ± 0.46 | 58.6 ± 0.22 | 60.28 ± 0.41 | 91.85 ± 0.24 |
| RFRS | 56.37 ± 0.58 | 58.41 ± 0.47 | 61.5 ± 0.5 | 93.59 ± 0.69 |
| SVM | 39.6 | 39.62 | 40.38 | 75.62 |
| AdaBoost | 42.32 ± 0.07 | 42.59 ± 0.36 | 28.95 ± 0.96 | 48.03 ± 0.04 |
| GB | 42.17 ± 0.35 | 42.21 ± 2.05 | 40.29 ± 0.17 | 49.83 ± 0.28 |
| XGB | 56.69 ± 0.18 | 58.05 ± 0.13 | 58.45 ± 0.83 | 89.29 ± 1.59 |
| SimpleANN | 40.56 ± 0.65 | 40.6 ± 0.56 | 41.22 ± 0.25 | 87.68 ± 0.94 |
| DropoutANN | 40.98 ± 0.05 | 40.35 ± 0.5 | 39.4 ± 0.82 | 91.24 ± 0.75 |
| ComplexANN | 36.95 ± 1.84 | 38.47 ± 0.54 | 36.87 ± 1.99 | 88.5 ± 1.68 |
| Model | SIDIC Dataset | | | |
| | 2000 Samples | 3000 Samples | 5000 Samples | 8500 Samples |
| CNN IC | 88.75 ± 0.4 | 91.2 ± 0.25 | 95.88 ± 0.22 | 98.72 ± 0.17 |

From the models trained on numerical data, the best performer was RFRS with a mean accuracy of 93.59. Although, the best model overall was CNN IC when trained on 8500 samples, reaching a staggering mean accuracy of 98.72. The worst model, besides NB was AdaBoost when trained on 5000 samples with a mean accuracy of 28.95. Ignoring the advantage of better performance, training RFRS took 4 hours, compared to the time it took CNN IC, which was 10 times longer. RFRS also proved to be more lightweight, and might be the best option when computational cost, RAM and time are considered of high importance.

## 5.2 Analysis of Results

As we hypothesized earlier, as the sample size increases so does accuracy. This seems to be true for almost all results. Even though the increase in accuracy, when moving from 2000 to 3000 and 5000 samples is notable for most models, there seems to be a sudden very significant jump when moving from 5000 to 8500 samples for training, affecting all models. As of now, we cannot elaborate why this is the exact case, because it is very hard to do so. Although, we can propose a hypothesis. One relevant study that elaborates on the phenomenon of a significant increase in model accuracy with an increase in sample size is by Morgan et al. in 2003 [47]. They explored how decision-tree based models' accuracy improves as the sample size increases. Their findings suggest that model accuracy improves at a decreasing rate with increasing sample size, but substantial improvements can occur as sample sizes reach certain thresholds, very similar to our observation from 5000 to 8500 samples. This is most likely the case. They also discuss using a power curve to fit accuracy estimates across various sample sizes, which could be a useful approach for understanding the dynamics in our data [47]. A power curve is a type of nonlinear mathematical model used to describe the relationship between two variables, in this case, sample size and model accuracy. The idea is to represent the accuracy of a model as a function of the size of the data set used to train the model. This behavior is modeled by fitting a power curve to the accuracy estimates across our various sample sizes. It can be modelled generically by using the following formula:

$$y = a * x^b$$

where $b$ and $a$ are constants, respectively, the exponent of the power law, and the width factor of the scaling relationship. Using this formula, we can fit the function to our sample sizes and predict accuracies and analyze the curve changes. See Figure 34 below.

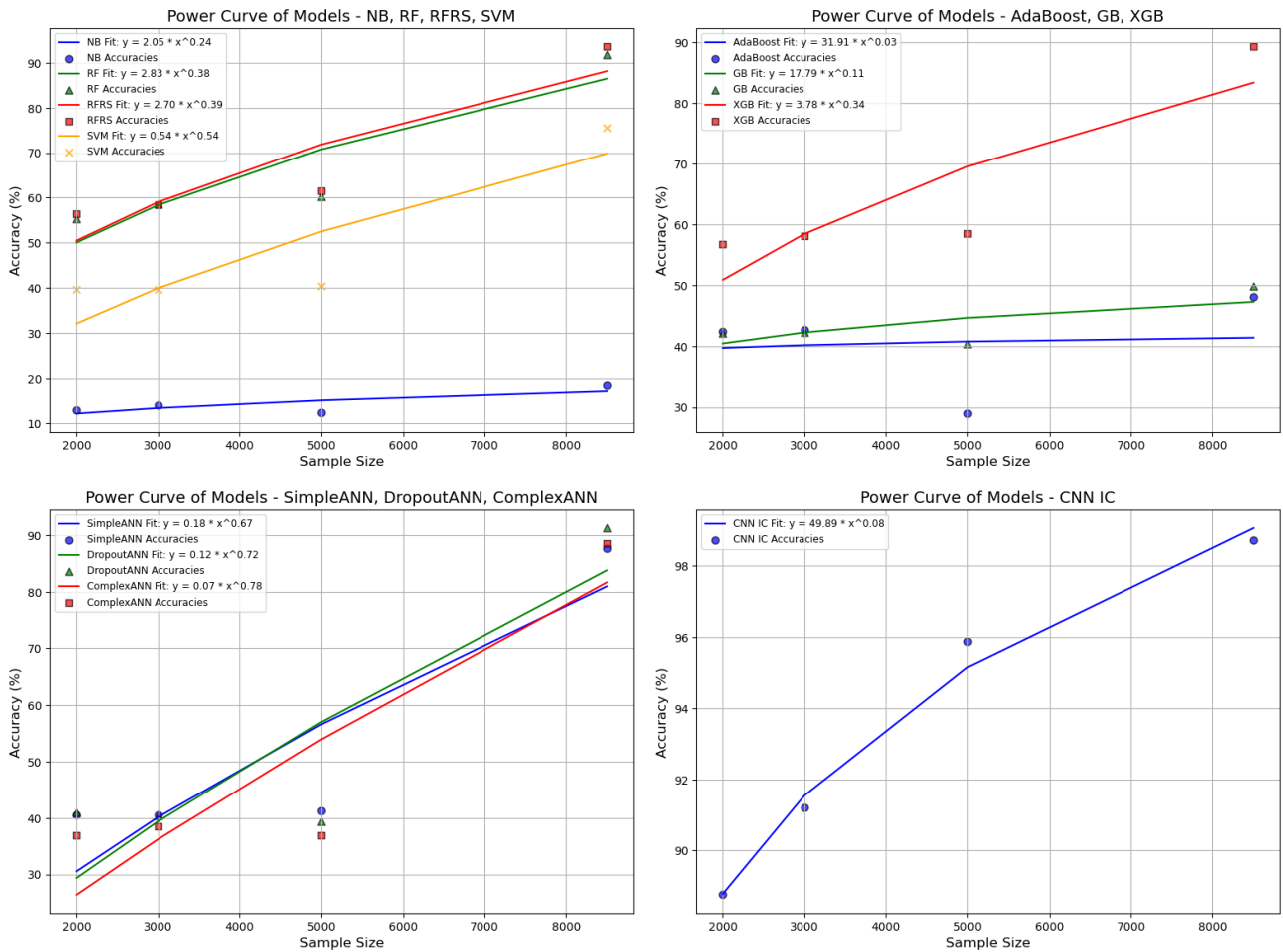

*Figure 34: Power curves of all our models, including the actual accuracy data points as scatter dots for comparison.*

Based on the curve shape, most models overperform for 2000 samples and 8500 samples, while underperforming for 5000 samples. The performance prediction seems to be very accurate for 3000 samples, as seen in the overlapping of the scatter data and the curve line. Conversely, for the case of the CNN IC model, it does the exact opposite, only outperforming on 5000 samples and underperforming on all other sample sizes, even 8500. We can clearly see the importance of data size in the tasks of MIR.

# 6 Dual-Input NN Instrument Classifier

In our pursuit of developing a superior AI classifier for musical instrument classification, we also introduce an innovative neural network architecture that utilizes a dual-input system. This combined approach leverages the strengths of both the NuFDIC and the SIDIC, combining diverse data modalities to enhance classification accuracy. By feeding the network with both numerical audio features and spectrogram images, the model can capture a more comprehensive understanding of the audio signals. The model input begins with an image input layer. This image stream passes through a series of convolutional layers: initially, a 32-filter layer followed by a 64-filter layer, both with 3x3 kernel sizes and ReLU activation functions. After the convolutional layers, a max-pooling layer with a 2x2 pool is applied. Simultaneously, the model processes numerical input. This data passes through two dense layers with 64 and 32 neurons respectively, each using ReLU activation. These two data streams are then concatenated, combining the flattened output of the image processing path with the output from the numerical data path, thereby merging the features extracted from both sources. This concatenated vector feeds into a series of dense layers, further enabling the model to learn more complex patterns from the combined feature set. This dual-input system allows the neural network to concurrently process and learn from the distinct but complementary information present in both datasets, potentially addressing limitations faced by the single-input models. Model architecture is given in Figure 35. The experimental setup involved training this hybrid model on both datasets, aiming to explore how such a synergistic approach could outperform models using only one type of data input. Due to limitations regarding time and cost, we only trained the model on 2000 samples for both datasets. A promising accuracy of 0.948 was reported within only 20 epochs of training.

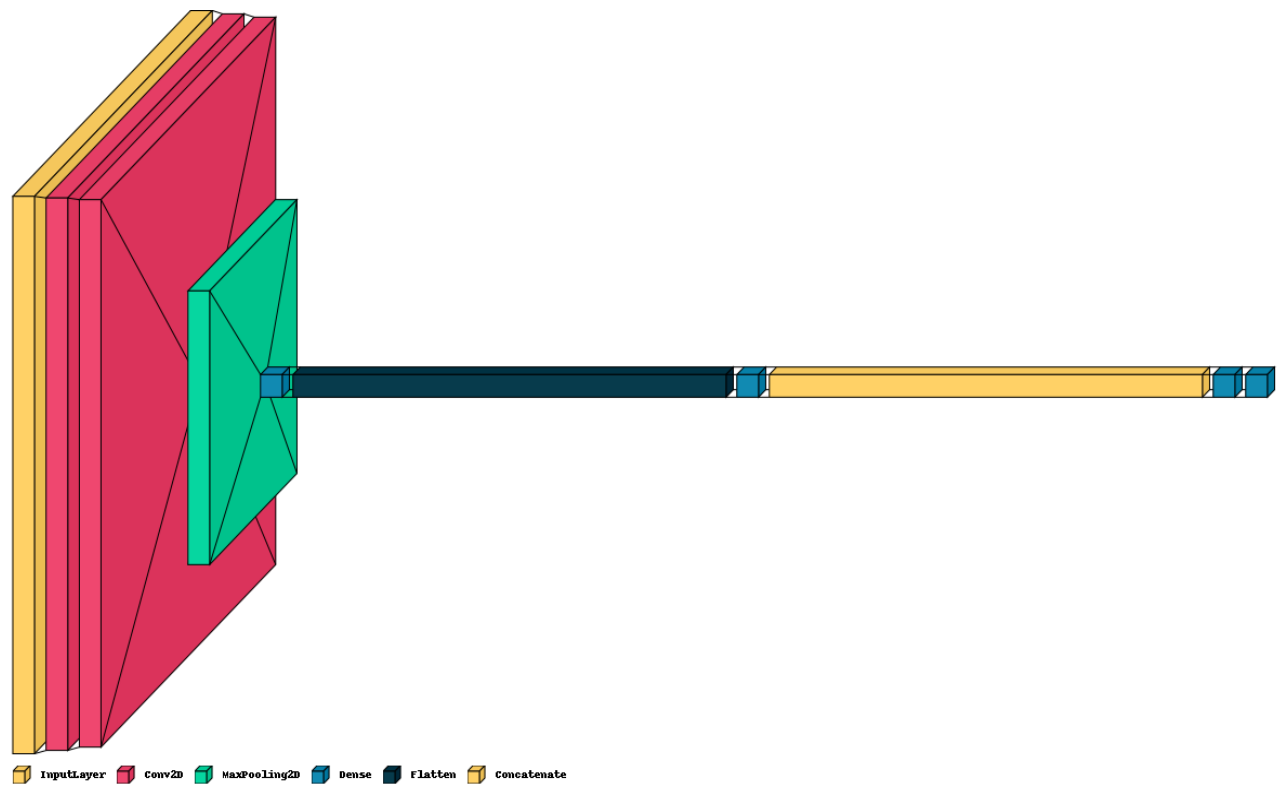

*Figure 35: Overview of our Dual-Input NN model for instrument classification.*

# 7    Conclusions and Future Work

This study aimed to evaluate the efficacy of various ML techniques in the classification of musical instruments. A broad range of methods, including traditional ML algorithms, boosting techniques, and DL models, were utilized to analyze audio data from the Nsynth dataset. Among traditional ML models, RFRS achieved the highest accuracy, particularly when trained on larger datasets. DL models, particularly the CNN IC, outperformed all other models. This underscores the advantage of using image-based data for complex classification tasks in MIR. The ANN models used in this study provided insights into how varying architectures and regularization techniques can impact the performance of DL systems in this task. Boosting algorithms showed varied performance, with XGB performing substantially better than AdaBoost and standard GB, highlighting the efficiency of modern ensemble methods that incorporate regularization and optimization techniques. There was a noticeable trend across all models, where increasing the sample size generally led to better model performance. This was particularly evident when moving from 5000 to 8500 samples, suggesting a threshold effect where larger datasets might be necessary to capture the complex variability in musical sounds adequately. Across several models, consistent misclassification between certain classes namely keyboard, mallet, and guitar were observed. This could be attributed to the acoustic similarities between these instruments, which poses a challenge for the models to distinguish based on the features extracted.

Regarding future work, the development of an optimal dual-input model may be competitive, as we explicitly tested. The testing in this area was limited, hence there is a lot of room for improvement through more testing. For audio data, incorporating temporal dynamics directly into ANN models using architectures like RNN or LSTM could potentially improve classification accuracy by capturing time-dependent features of audio signals. Thus, testing on various RNN models could still be implemented. Applying the Fourier Transform to audio signals can help extract detailed frequency-based features that are essential for identifying and differentiating between musical instruments. FFT generated images may prove to be useful as well. The Gramian Angular Field (GAF) is a novel transformation technique, introduced in 2015 that converts time series data into a matrix format, which can then be visualized as an image. This approach opened new possibilities for analyzing temporal data using methods traditionally reserved for image processing [48]. For audio signals, GAFs can highlight the cyclic patterns that are characteristic of different musical instruments. Thus, utilizing GAF images as data for CNNs may lead to very promising results. Lastly, developing models that can perform classification in real-time could be valuable for applications in digital music production and live performance settings.

**The source code for all the experiments conducted in this study is publicly available at: https://github.com/JoanikijChulev/AI-Instrument-Classification.**

# References

[1] B. S. Manjunath and W. Y. Ma, "Texture features for browsing and retrieval of image data," *IEEE Trans Pattern Anal Mach Intell*, vol. 18, no. 8, pp. 837–842, 1996, doi: 10.1109/34.531803.

[2] M. Mitrevski, "Natural Language Processing on iOS," in *Developing Conversational Interfaces for iOS*, Berkeley, CA: Apress, 2018, pp. 165–185. doi: 10.1007/978-1-4842-3396-2_6.

[3] J. F. Gemmeke *et al.*, "Audio Set: An ontology and human-labeled dataset for audio events," in *2017 IEEE International Conference on Acoustics, Speech and Signal Processing (ICASSP)*, IEEE, Mar. 2017, pp. 776–780. doi: 10.1109/ICASSP.2017.7952261.

[4] B. W. Schuller, *Intelligent Audio Analysis*. Berlin, Heidelberg: Springer Berlin Heidelberg, 2013. doi: 10.1007/978-3-642-36806-6.

[5] S. Pfeiffer, S. Fischer, and W. Effelsberg, "Automatic audio content analysis," in *Proceedings of the fourth ACM international conference on Multimedia - MULTIMEDIA '96*, New York, New York, USA: ACM Press, 1996, pp. 21–30. doi: 10.1145/244130.244139.

[6] K. Ohtani, K. Niwa, and K. Takeda, "AI framework to arrange audio objects according to listener preferences," *Journal of the Acoustical Society of America*, vol. 140, no. 4_Supplement, pp. 3060–3060, Oct. 2016, doi: 10.1121/1.4969523.

[7] T. Giannakopoulos, "pyAudioAnalysis: An Open-Source Python Library for Audio Signal Analysis," *PLoS One*, vol. 10, no. 12, p. e0144610, Dec. 2015, doi: 10.1371/journal.pone.0144610.

[8] D. M. Chandwadkar and M. S. Sutaone, "Selecting Proper Features and Classifiers for Accurate Identification of Musical Instruments," *Int J Mach Learn Comput*, pp. 172–175, 2013, doi: 10.7763/ijmlc.2013.v3.296.

[9] R. Profeta and G. Schuller, "End-to-End Learning for Musical Instruments Classification," in *2021 55th Asilomar Conference on Signals, Systems, and Computers*, IEEE, Oct. 2021, pp. 1607–1611. doi: 10.1109/IEEECONF53345.2021.9723181.

[10] S. Prabavathy, "Classification of Musical Instruments Sound Using Pre-Trained Model with Machine Learning Techniques," *Asian Journal of Electrical Sciences*, vol. 9, no. 1, pp. 45–48, May 2020, doi: 10.51983/ajes-2020.9.1.2369.

[11] A. A. Harryanto, K. Gunawan, R. Nagano, and R. Sutoyo, "Music Classification Model Development Based on Audio Recognition using Transformer Model," in *2022 3rd International Conference on Artificial Intelligence and Data Sciences (AiDAS)*, IEEE, Sep. 2022, pp. 258–263. doi: 10.1109/AiDAS56890.2022.9918787.

[12] L. Zhang, I. Saleh, S. Thapaliya, J. Louie, J. Figueroa-Hernandez, and H. Ji, "An Empirical Evaluation of Machine Learning Approaches for Species Identification through Bioacoustics," in *2017 International Conference on Computational Science and Computational Intelligence (CSCI)*, IEEE, Dec. 2017, pp. 489–494. doi: 10.1109/CSCI.2017.82.

[13] N. Kawwa, "NadimKawwa/NSynth: Instrument classification on the NSynth dataset using supervised learning and CNNs." Accessed: Apr. 17, 2024. [Online]. Available: https://github.com/NadimKawwa/NSynth

[14] J. Engel *et al.*, "Neural Audio Synthesis of Musical Notes with WaveNet Autoencoders," Apr. 2017, [Online]. Available: http://arxiv.org/abs/1704.01279

[15] I. Shimoda and S. Fujioka, "Percussive musical instrument," *J Acoust Soc Am*, vol. 85, no. 4, pp. 1813–1813, Apr. 1989, doi: 10.1121/1.397877.

[16] H. Tachibana, N. Ono, H. Kameoka, and S. Sagayama, "Harmonic/Percussive Sound Separation Based on Anisotropic Smoothness of Spectrograms," *IEEE/ACM Trans Audio Speech Lang Process*, vol. 22, no. 12, pp. 2059–2073, Dec. 2014, doi: 10.1109/TASLP.2014.2351131.

[17] A. Kumar Shah and A. Nepal, "Chroma Feature Extraction." [Online]. Available: https://www.researchgate.net/publication/330796993

[18] A. Ghosal, S. Dutta, and D. Banerjee, "Stratification of String Instruments Using Chroma-Based Features," 2019, pp. 181–191. doi: 10.1007/978-981-13-1951-8_17.

[19] N. Agera, S. Chapaneri, and D. Jayaswal, "Exploring Textural Features for Automatic Music Genre Classification," in *2015 International Conference on Computing Communication Control and Automation*, IEEE, Feb. 2015, pp. 822–826. doi: 10.1109/ICCUBEA.2015.164.

[20] N. Rodin, D. Pinčić, K. Lenac, and D. Sušanj, "The Comparison of Different Feature Extraction Methods in Musical Instrument Classification," in *2023 46th MIPRO ICT and Electronics Convention (MIPRO)*, IEEE, May 2023, pp. 1136–1141. doi: 10.23919/MIPRO57284.2023.10159952.

[21] H. M. Fayek, "Speech Processing for Machine Learning: Filter banks, Mel-Frequency Cepstral Coefficients (MFCCs) and What's In-Between," https://haythamfayek.com/2016/04/21/speech-processing-for-machine-learning.html.

[22] E. L. Salomons and P. J. M. Havinga, "A survey on the feasibility of sound classification on wireless sensor nodes," *Sensors (Switzerland)*, vol. 15, no. 4, pp. 7462–7498, Mar. 2015, doi: 10.3390/s150407462.

[23] W. Shen, D. Tu, Y. Yin, and J. Bao, "A new fusion feature based on convolutional neural network for pig cough recognition in field situations," *Information Processing in Agriculture*, vol. 8, no. 4, pp. 573–580, Dec. 2021, doi: 10.1016/J.INPA.2020.11.003.

[24] M. S. Nagawade and V. R. Ratnaparkhe, "Musical instrument identification using MFCC," in *2017 2nd IEEE International Conference on Recent Trends in Electronics, Information & Communication Technology (RTEICT)*, IEEE, May 2017, pp. 2198–2202. doi: 10.1109/RTEICT.2017.8256990.

[25] S. Prabavathy*, V. Rathikarani, and P. Dhanalakshmi, "Classification of Musical Instruments using SVM and KNN," *International Journal of Innovative Technology and Exploring Engineering*, vol. 9, no. 7, pp. 1186–1190, May 2020, doi: 10.35940/ijitee.G5836.059720.

[26] Dan-Ning Jiang, Lie Lu, Hong-Jiang Zhang, Jian-Hua Tao, and Lian-Hong Cai, "Music type classification by spectral contrast feature," in *Proceedings. IEEE International Conference on Multimedia and Expo*, IEEE, pp. 113–116. doi: 10.1109/ICME.2002.1035731.

[27] S. Jin-Soo, "Study on the Performance of Spectral Contrast MFCC for Musical Genre Classification." Accessed: Apr. 17, 2024. [Online]. Available: https://www.researchgate.net/publication/264169888_Study_on_the_Performance_of_Spectral_Contrast_MFCC_for_Musical_Genre_Classification

[28] B. McFee *et al.*, "librosa: Audio and Music Signal Analysis in Python," *Proceedings of the 14th Python in Science Conference*, pp. 18–24, 2015, doi: 10.25080/MAJORA-7B98E3ED-003.

[29] H. He and E. A. Garcia, "Learning from imbalanced data," *IEEE Trans Knowl Data Eng*, vol. 21, no. 9, pp. 1263–1284, Sep. 2009, doi: 10.1109/TKDE.2008.239.

[30] M. Banko and E. Brill, "Scaling to Very Very Large Corpora for Natural Language Disambiguation," pp. 26–33, 2001, doi: 10.3115/1073012.1073017.

[31] C. Sun, A. Shrivastava, S. Singh, and A. Gupta, "Revisiting Unreasonable Effectiveness of Data in Deep Learning Era," *Proceedings of the IEEE International Conference on Computer Vision*, vol. 2017-October, pp. 843–852, Jul. 2017, doi: 10.1109/ICCV.2017.97.

[32] A. Natsiou and S. O'Leary, "A sinusoidal signal reconstruction method for the inversion of the mel-spectrogram," in *2021 IEEE International Symposium on Multimedia (ISM)*, IEEE, Nov. 2021, pp. 245–248. doi: 10.1109/ISM52913.2021.00048.

[33] Y. M. G. Costa, L. S. Oliveira, and C. N. Silla, "An evaluation of Convolutional Neural Networks for music classification using spectrograms," *Appl Soft Comput*, vol. 52, pp. 28–38, Mar. 2017, doi: 10.1016/j.asoc.2016.12.024.

[34] Z. Fu, G. Lu, K. M. Ting, and D. Zhang, "Learning Naive Bayes Classifiers for Music Classification and Retrieval," in *2010 20th International Conference on Pattern Recognition*, IEEE, Aug. 2010, pp. 4589–4592. doi: 10.1109/ICPR.2010.1121.

[35] Y. Zhang and D. LV, "Selected Features for Classifying Environmental Audio Data with Random Forest," *The Open Automation and Control Systems Journal*, vol. 7, no. 1, pp. 135–142, Mar. 2015, doi: 10.2174/1874444301507010135.

[36] Y. Zhu, Z. Ming, and Q. Huang, "Automatic Audio Genre Classification Based on Support Vector Machine," in *Third International Conference on Natural Computation (ICNC 2007)*, IEEE, 2007, pp. 517–521. doi: 10.1109/ICNC.2007.277.

[37] J. Bergstra, N. Casagrande, D. Erhan, D. Eck, and B. Kégl, "Aggregate features and ADABOOST for music classification," *Mach Learn*, vol. 65, no. 2–3, pp. 473–484, Dec. 2006, doi: 10.1007/s10994-006-9019-7.

[38] P. Bahad and P. Saxena, "Study of AdaBoost and Gradient Boosting Algorithms for Predictive Analytics," pp. 235–244, 2020, doi: 10.1007/978-981-15-0633-8_22.

[39] Y. Liu, Y. Yin, Q. Zhu, and W. Cui, "Musical Instrument Recognition by XGBoost Combining Feature Fusion," Jun. 2022, Accessed: Apr. 24, 2024. [Online]. Available: https://arxiv.org/abs/2206.00901v1

[40] S. K. Mahanta, A. F. U. Rahman Khilji, and P. Pakray, "Deep Neural Network for Musical Instrument Recognition Using MFCCs," *Computación y Sistemas*, vol. 25, no. 2, pp. 351–360, May 2021, doi: 10.13053/cys-25-2-3946.

[41] J. R. De Gruijl and M. A. Wiering, "Musical Instrument Classification using Democratic Liquid State Machines."

[42] Y. Su, "Instrument Classification Using Different Machine Learning and Deep Learning Methods," 2023.

[43] "Confusion Matrix in Machine learning." Accessed: Apr. 26, 2024. [Online]. Available: https://www.analyticsvidhya.com/blog/2020/04/confusion-matrix-machine-learning/

[44] D. Krstinić, M. Braović, L. Šerić, and D. Božić-Štulić, "Multi-label Classifier Performance Evaluation with Confusion Matrix," pp. 01–14, Jun. 2020, doi: 10.5121/CSIT.2020.100801.

[45] T. Fawcett, "An introduction to ROC analysis," *Pattern Recognit Lett*, vol. 27, no. 8, pp. 861–874, Jun. 2006, doi: 10.1016/J.PATREC.2005.10.010.

[46] D. P. Kingma and J. Ba, "Adam: A Method for Stochastic Optimization," Dec. 2014, [Online]. Available: http://arxiv.org/abs/1412.6980

[47] J. Morgan, R. Dougherty, A. Hilchie, and B. Carey, "Sample Size and Modeling Accuracy with Decision Tree Based Data Mining Tools SAMPLE SIZE AND MODELING ACCURACY WITH DECISION-TREE BASED DATA MINING TOOLS," 2003. Accessed: May 02, 2024. [Online]. Available: https://www.researchgate.net/publication/238689433_Sample_Size_and_Modeling_Accuracy_with_Decision_Tree_Based_Data_Mining_Tools

[48] Z. Wang and T. Oates, "Imaging Time-Series to Improve Classification and Imputation".